\documentclass[preprint2,tighten]{aastex62_reep}

\usepackage{epstopdf}
\usepackage{subfigure}  
\usepackage{amsmath}

\usepackage{longtable}

\usepackage{fancyhdr}   
\fancypagestyle{specialfooter}{%
 \fancyhf{}
 
 \fancyfoot[C]{Distribution Statement A.  Approved for public release.  Distribution unlimited.}  }

\shorttitle{Electron Beams Cannot Directly Produce Coronal Rain}
\shortauthors{Reep, Antolin, \& Bradshaw}

\begin{document}

\title{Electron Beams Cannot Directly Produce Coronal Rain}

\author[0000-0003-4739-1152]{Jeffrey W. Reep}
\affiliation{Space Science Division, Naval Research Laboratory, Washington, DC 20375, USA; \href{mailto:jeffrey.reep@nrl.navy.mil}{jeffrey.reep@nrl.navy.mil}}

\author[0000-0003-1529-4681]{Patrick Antolin}
\affiliation{Department of Mathematics, Physics, and Electrical Engineering, Northumbria University, Newcastle upon Tyne, NE1 8ST, UK}

\author[0000-0002-3300-6041]{Stephen J. Bradshaw}
\affiliation{Department of Physics \& Astronomy, Rice University, Houston, TX 77005, USA}

\begin{abstract}
Coronal rain is ubiquitous in flare loops, forming shortly after the onset of the solar flare.  Rain is thought to be caused by a thermal instability, a localized runaway cooling of material in the corona.  The models that demonstrate this require extremely long duration heating on the order of the radiative cooling time, localized near the footpoints of the loops.  In flares, electron beams are thought to be the primary energy transport mechanism, driving strong footpoint heating during the impulsive phase that causes evaporation, filling and heating flare loops.  Electron beams, however, do not act for a long period of time, and even supposing that they did, their heating would not remain localized at the footpoints.  With a series of numerical experiments, we show directly that these two issues mean that electron beams are incapable of causing the formation of rain in flare loops.  This result suggests that either there is another mechanism acting in flare loops responsible for rain, or that the modeling of the cooling of flare loops is somehow deficient.  To adequately describe flares, the standard model must address this issue to account for the presence of coronal rain.
\end{abstract}

\keywords{Sun: atmosphere, Sun: flares, Sun: corona}

\nopagebreak

\section{Introduction}
\label{sec:intro}
\thispagestyle{specialfooter}   

Coronal rain forms in the gradual phase of solar flares, beginning around the end of the impulsive phase.  Appearing as blobs of plasma that fall down coronal loops, rain is seen regularly in observations of the corona (\textit{e.g.} \citealt{foukal1978}) and in flare loops at cool temperatures such as H$\alpha$ (\textit{e.g.} \citealt{jing2016}), \ion{He}{2} 304\,\AA\ \citep{scullion2016}, and IRIS slit-jaw images \citep{lacatus2017}.  \citet{jing2016} showed, importantly, that the impact of the rain in the chromosphere followed exactly the same path as the flare ribbon and had the same size scale as the ribbon, demonstrating a tight cause and effect between the energy release and formation of rain.  

In the quiescent, non-flaring context, rain is thought to be caused by a thermal instability in the plasma \citep{parker1953,field1965} due to localized runaway cooling, a key feature of thermal non-equilibrium (TNE; see discussion by \citealt{antolin2020} and \citealt{klimchuk2019b}).  Long duration, quasi-steady heating localized near the footpoints of loops causes cycles of mass transfer between the corona and chromosphere that prevent the formation of a stable equilibrium \citep{kuin1982,antiochos1999}.  At the end of the cooling state of that cycle, when the temperature is nearly constant across the corona and thermal conduction therefore is negligible, the loop is in a critical state of equilibrium, allowing a thermal instability to kick in locally due to a perturbation, where radiation ($\propto n^{2}$) causes a runaway cooling that forms a so-called coronal condensation.  The high-density, low-temperature condensation appears in chromospheric and transition region wavelengths as a localized brightening, which falls along the field line towards the chromosphere \citep{antolin2012,oliver2014}.

The formation of TNE has been extensively studied with numerical modeling.  The comprehensive work of \citet{froment2018} recently showed that there is a specific parameter space within which TNE occurs, which depends strongly on the heating rate, heating location, and asymmetry in the heating.  The observational signature of the TNE cycles in extreme ultraviolet (EUV) passbands is thought to be the recently discovered long-period intensity pulsations (\textit{e.g.} \citealt{froment2015,auchere2018}).  The work of \citet{mikic2013} showed that TNE and the formation of coronal condensations also depends strongly on the geometry of the loop, both in terms of cross-sectional area expansion and a general loop asymmetry.  The requirement that the heating be steady over a long period of time, however, suggests an issue: rain is observed in flares \citep{jing2016} and perhaps in non-flaring impulsive events as well \citep{kohutova2019}.  

Solar flares are driven by magnetic reconnection, where magnetic energy is rapidly converted into thermal energy, kinetic energy, and wave motions \citep{carmichael1964,sturrock1966,kopp1976}.  The sudden release of energy leads to substantial heating of the plasma and the characteristic eruptions and brightenings (\textit{e.g.} \citealt{warren2018}).  The conversion of magnetic energy to kinetic energy, in particular the acceleration of electrons, is known to be a crucial component of energy transport in flares \citep{holman2011}.  Electrons are accelerated in the corona to energies exceeding 10 keV, perhaps up to hundreds of keV \citep{holman2003}, and stream towards the lower atmosphere, where they deposit energy through collisions with the ambient plasma \citep{emslie1978}.  This energy deposition then drives an increase in pressure, leading to an expansion of plasma and ablation of material back into the corona (termed chromospheric evaporation\footnote{Note that there is no change in state, so the word ``evaporation'' is technically incorrect.  The same is also true of coronal ``condensations''.}, \citealt{hirayama1974}), filling the flare loops with hot and dense material.  As a result, the intensities at X-ray and extreme ultraviolet (EUV) wavelengths brighten sharply, leading to the typical observed properties of flares \citep{fletcher2011}.  

The energy release in solar flares is by its very nature impulsive, suggesting that it may be fundamentally incompatible with the high-frequency heating in the TNE picture.  The duration of flares in the soft X-rays occurs on a log-normal distribution, with median full-width-at-half-maximum durations of approximately 10 minutes \citep{reepknizhnik2019}.  This means that the vast majority of flares are not heated long enough to produce the cycles characteristic of TNE, and it is not clear how a thermal instability would then form (see also the discussion in \citealt{antiochos1980}).  

To date, there have been no attempts to model the formation of coronal rain in flares with hydrodynamic simulations.  While the observations are unambiguous that rain forms in flare loops, hydrodynamic modeling has generally focused on the impulsive phase, and failed entirely to account for rain.  In this work, we test directly whether standard heating by an electron beam can produce coronal rain, surprisingly finding an absence of rain in the cooling phase of the model.  This leads us to two distinct possibilities: either electron beam heating is inadequate to explain the formation coronal rain in flares, or the cooling of the plasma is missing a crucial process.  We discuss both of these, and give arguments that both are problematic.  We conclude that there is a startling weakness in our understanding of flares: the standard model does not account for coronal rain.

\section{Electron Beams}
\label{sec:beams}

Non-thermal electrons are stopped collisionally as they traverse a flaring loop.  Due to the sharp increase in density as one descends from the corona into the chromosphere, the bulk of the energy deposition occurs in the chromosphere initially.  The column density from the injection site $N = \int_{z_0}^{z} n(z')\ dz'$ therefore determines the average stopping depth for an injected electron spectrum.  For a sharp cut-off, the mean stopping depth is approximately $N_{c} = \frac{E_{c}^{2}}{6 \pi e^{4} \Lambda}$ \citep{nagai1984}, where $E_{c}$ is the low-energy cut-off of the electron beam, $e$ is the electron charge, and $\Lambda$ is the Coulomb logarithm.  As the chromosphere is heated and plasma ablates into the corona, the distance from the injection site to the mean stopping depth decreases.  If it were assumed that an electron beam lasts on the order of 10\,s, this is a negligible effect, and the bulk of the energy is deposited in the chromosphere.  If, however, we assume that the electron beam acts upon one loop for an extended period of time ($\approx$ minutes), then the energy deposition location propagates towards the injection site of the beam.  

\begin{figure*}
\includegraphics[width=0.495\linewidth]{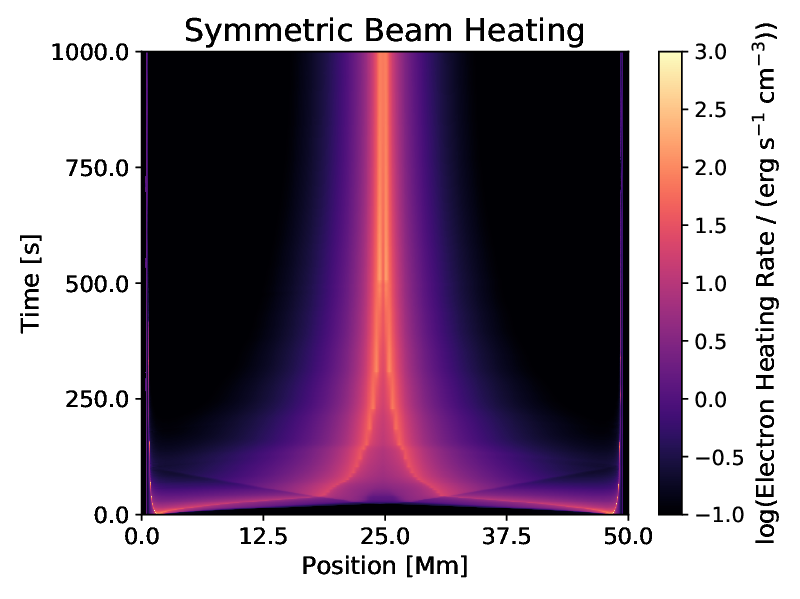}
\includegraphics[width=0.495\linewidth]{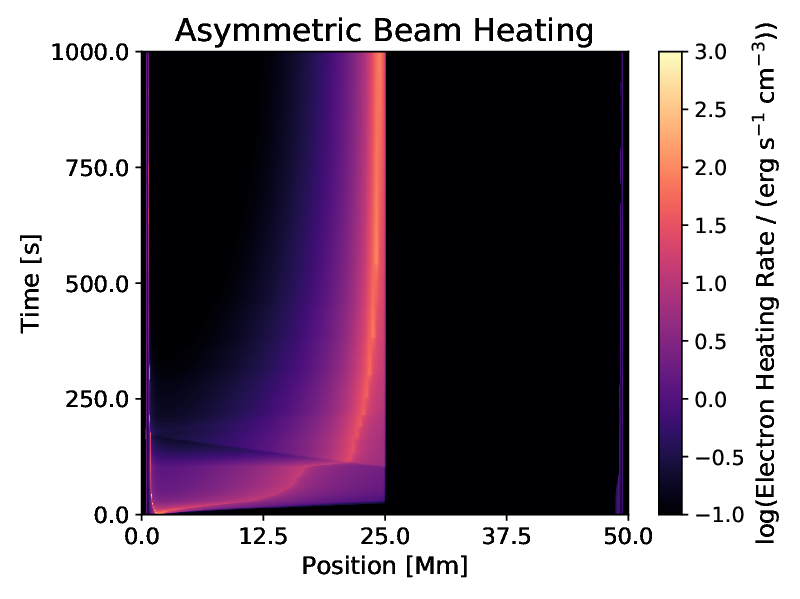}
\includegraphics[width=0.495\linewidth]{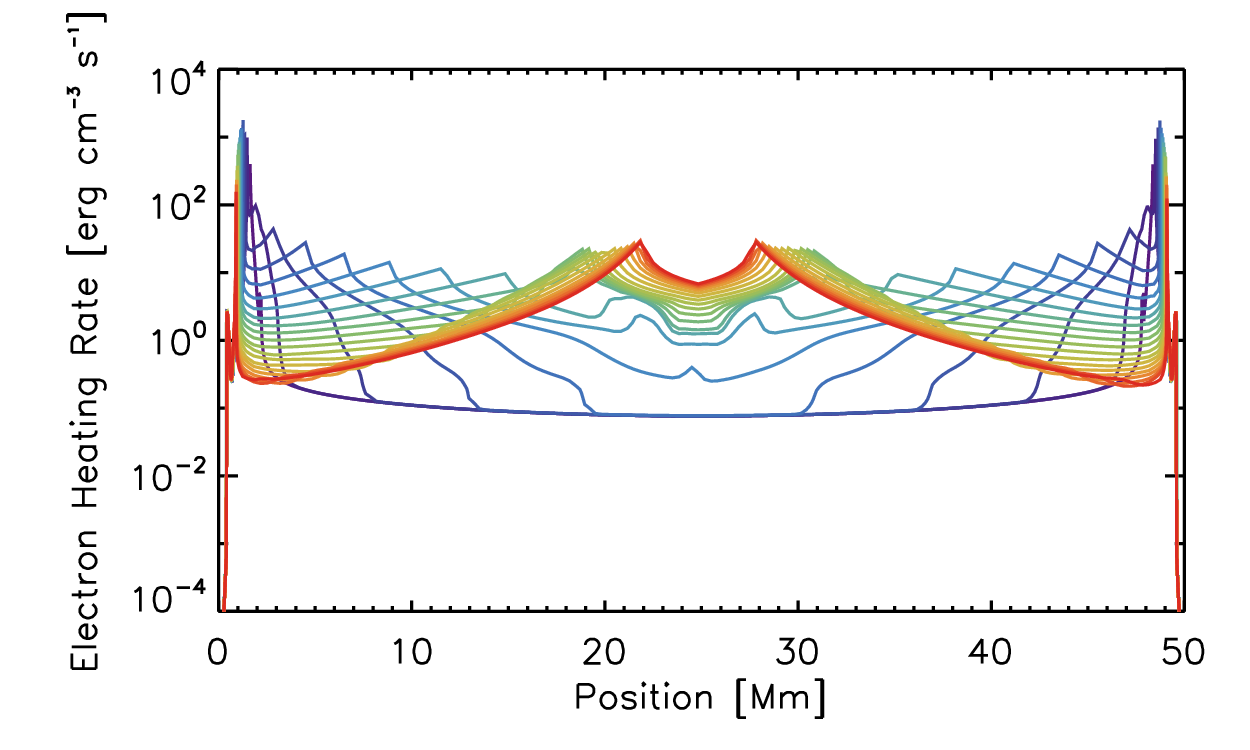}
\includegraphics[width=0.495\linewidth]{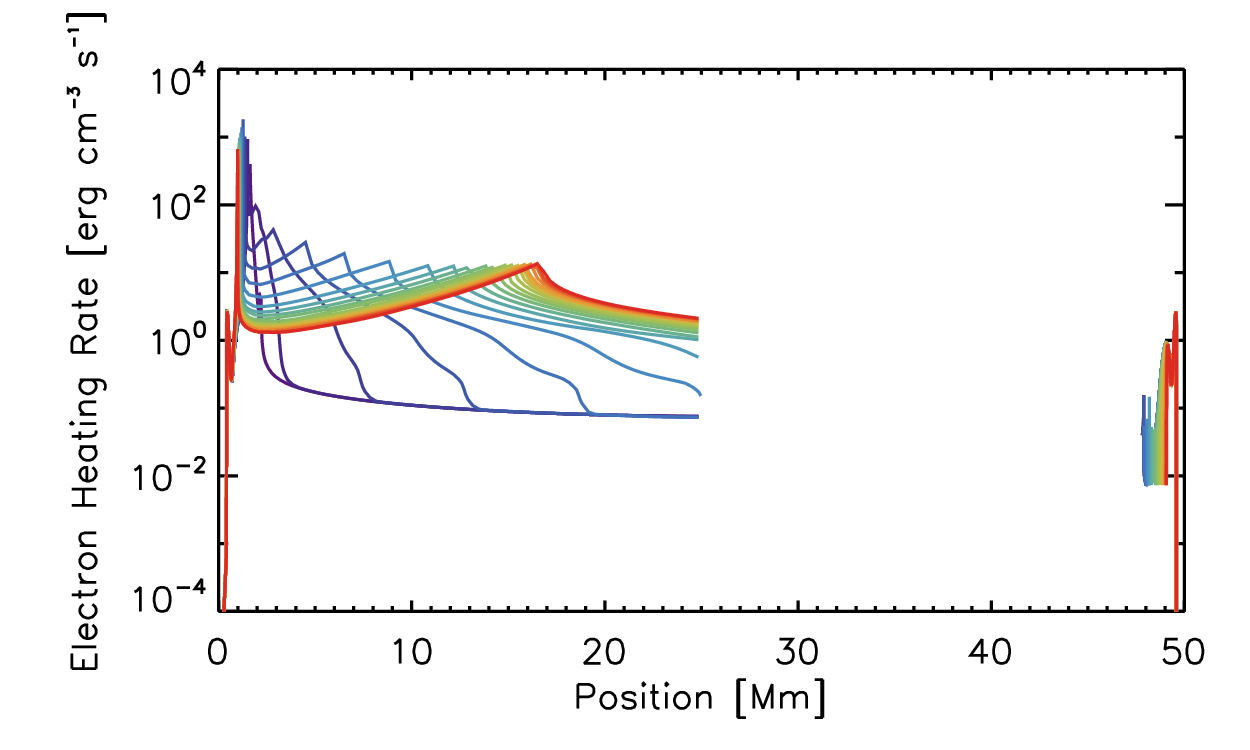}
\caption{The heating due to electron beams does not remain fixed at the footpoints.  The top plots show the change in the location and magnitude of energy deposition due to a long-lasting beam, as a function of position along the loop (x-axis) and time (y-axis), for a symmetric (left) and asymmetric beam (right).  The bottom plots show the same evolution on a line plot, with colors ranging from blue to red showing the evolution during the first 100 seconds of each simulation at a 5 second cadence.  While the bulk of the energy is initially deposited in the chromosphere, the location of maximal energy deposition quickly rises and propagates towards the injection site of the electron beam with time.}
\label{fig:heat_location}
\end{figure*}

For example, in Figure \ref{fig:heat_location}, we show the energy deposition location for an electron beam that is assumed to last indefinitely with for a case where the heating is symmetric to both legs of the loop (left) and the heating is asymmetric, heating only the left hand leg of the loop (right).  The top plots show the evolution along the loop (x-axis) as a function of time (y-axis), while the bottom plots show the same evolution on a line plot for the first 100 seconds of each simulation, where colors ranging from blue to red show the evolution at a 5 second cadence.  Initially, the heating is extremely strongly localized at the top of the chromosphere, but as evaporation fills the loop, the energy deposition in the corona rises sharply.  Regardless of the parameters of the beam (energy flux, cut-off, spectral index, asymmetry), the tendency is for the site of energy deposition to propagate towards the injection site of the electrons (see also \citealt{reep2015}).  This simple fact means that an electron beam cannot heat the footpoint of a single loop for an indefinite period of time.\footnote{This reasoning presupposes that the low-energy cut-off does not continuously and significantly increase while the beam acts, which would have the effect of causing the mean stopping depth to increase.  Such a scenario is not supported by any published observations, however.}  

Although other authors have not explicitly commented on the change in location of energy deposition for long-lasting beams, it can be seen in simulations performed with other codes in the published literature.  For example, Figure 1k of \citet{kerr2016} shows the evolution of the heating rate due to an electron beam with the RADYN code.  The heat deposition is initially localized strongly in the chromosphere, but as evaporation fills the loop, the heating rate in the corona rises along with the evaporation front.  Similar behavior is seen in the evolution of heating due to electron beams in the Flarix code, as shown in \textit{e.g.} Figure 2 of \citet{moravec2016}.  As evaporation drives the flow of plasma into the corona, the mean stopping depth is reduced and the energy deposition in the low corona begins to increase.  The vast majority of published studies of flare loop modeling, however, assume that the heating duration is of the order 10--20 seconds, and so this effect is not typically noticeable.

The natural question to ask then is how long electron beams do act upon a single loop.  Unfortunately, due to limited spatial resolution of hard X-ray (HXR) imagers, this duration cannot be directly measured.  For a flare as a whole, the hard X-ray burst lasts perhaps 5--10 minutes, suggesting an upper limit for the duration of a beam.  Spatially unresolved observations of bursty spikes in HXR light curves have been observed to occur at time-scales ranging from fractions of a second \citep{kiplinger1983,cheng2012} to tens of seconds \citep{lin1984}, with a wide variability.  This sub-structuring may suggest that the heating period is short.  The strong correlation between HXR and transition region line emission \citep{cheng1981,warren2001} suggests that even at the size of an \textit{IRIS} pixel (1/3 arcsec) there is sub-structuring at short time scales \citep{warren2016}.  However, multithreaded modeling of GOES and Yohkoh emission found improved consistency with observations when a long heating duration (200\,s) was assumed \citep{warren2006}.  A recent multithreaded modeling study \citep{reeppolito2018} found that the blue-shifts in \ion{Fe}{21} observed by \textit{IRIS} are most consistent with a distribution of heating durations, with mean value around 50--100\,s.  While there is no clear consensus yet, we consider it plausible as an upper limit that beams can act on a single loop for durations up to a few minutes, but not significantly longer.

There are two primary issues therefore that likely prevent electron beams from producing coronal rain or TNE:
\begin{enumerate}
    \item Electron beams do not last longer than a few minutes at most, while simulations that produce TNE consistently require durations at least an order magnitude longer.
    \item Even if they did last longer, electron beams do not consistently deposit their energy at the footpoints: the energy deposition propagates into the corona toward the apex (injection site) of the loop.
\end{enumerate}
This reasoning has led us to test directly whether electron beams, for some combination of parameters, can produce coronal rain.  

We use the HYDrodynamics and RADiation code (HYDRAD, \citealt{bradshaw2003}) to examine the capability of electron beams to produce coronal rain.  HYDRAD solves the hydrodynamic equations describing the conservation of mass, momentum, and energy for a two-fluid plasma along a full loop from the photosphere through the corona \citep{bradshaw2013}.  HYDRAD uses a full radiative loss calculation with CHIANTI \citep{dere2019}, as well as the prescription to optically thick losses in the chromosphere derived by \citet{carlsson2012}, thermal conduction with a flux limiting term, and the ability to solve for non-equilibrium ionization states.  Importantly, for this work, HYDRAD includes a magnetic expansion factor (equivalently, cross-sectional area expansion) from footpoint to corona, which has been shown to change the dynamics of a flaring loop \citep{emslie1992} and affect the likelihood of TNE \citep{mikic2013,klimchuk2019}.  

Electron beam heating has been previously implemented in HYDRAD \citep{reep2013,reep2016,reep2019}.  We use the heating function derived by \citet{emslie1978}, though a recent study suggested a modification to this to include the effect of diffusion in pitch angle, which effectively decreases the height at which electrons deposit their energy \citep{emslie2018}.  In this work, we assume an injected electron flux spectrum with a sharp low-energy cut-off:
\begin{equation}
    \mathfrak{F}(E_{0}, t) = \frac{F_{0}(t)}{E_{c}^{2}} (\delta - 2)         \Big(\frac{E_{0}}{E_{c}} \Big)^{-\delta}, \text{ for } E_{0} \geq           E_{c}  
\end{equation}
where $E_{0}$ (erg) is the energy of an electron at injection, $E_{c}$ is the so-called low-energy cut-off (erg), $F_{0}(t)$ is the energy flux of the beam (erg\,s$^{-1}$\,cm$^{-2}$), and $\delta$ is the spectral index.  We assume here that $E_{c}$ and $\delta$ do not vary with time.  

\begin{figure*}
\centering
\includegraphics[width=0.48\linewidth]{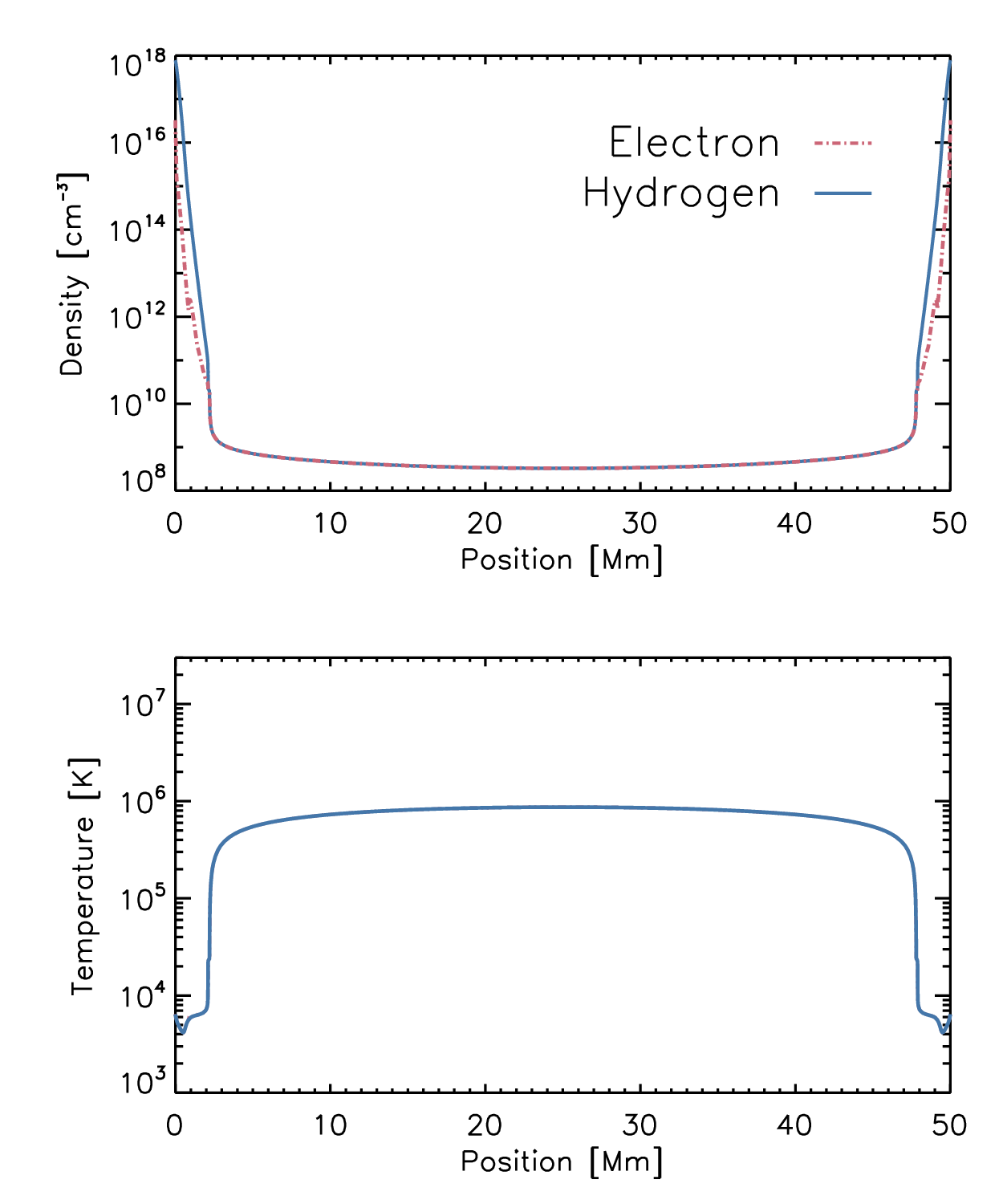}
\includegraphics[width=0.48\linewidth]{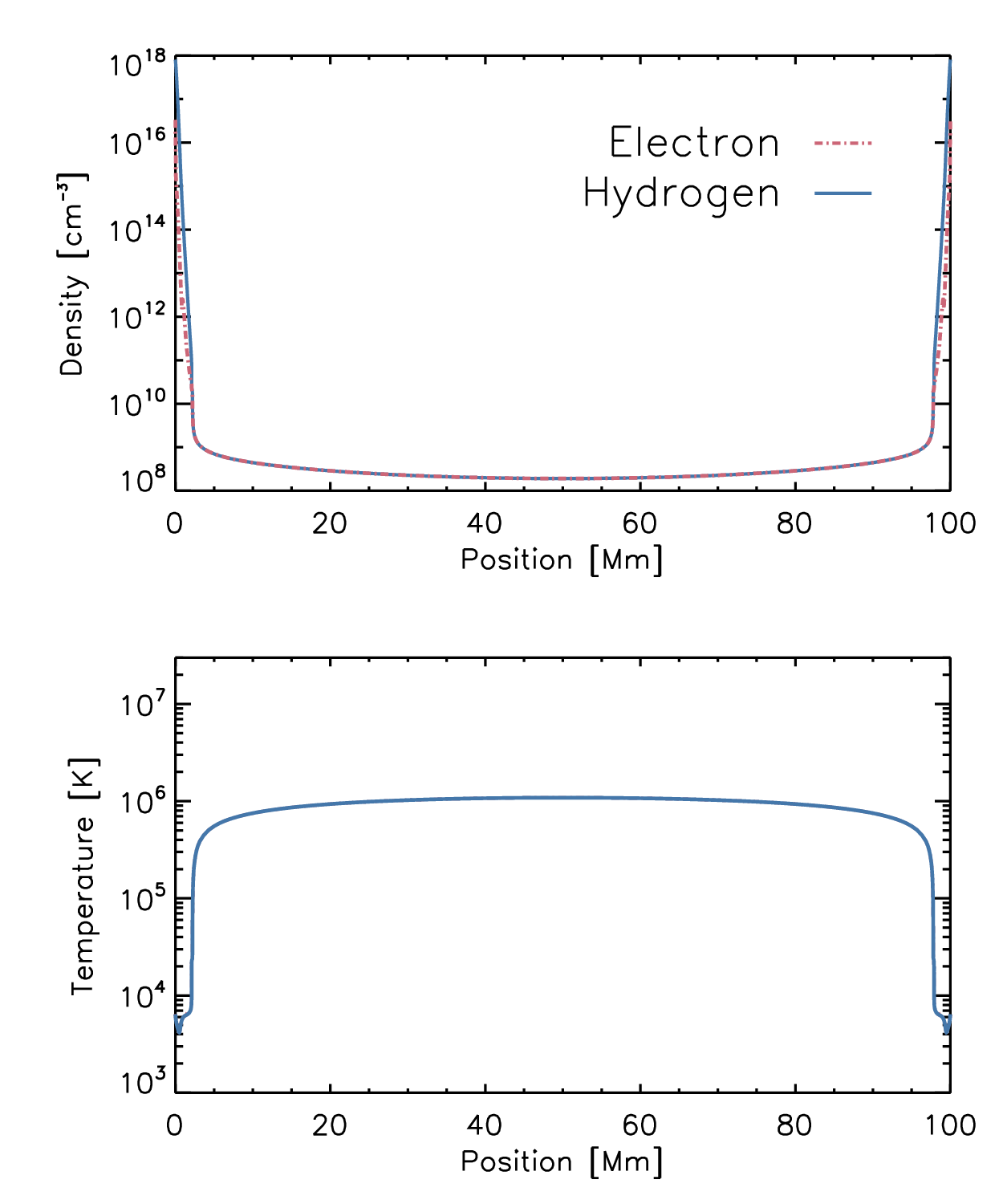}
\caption{The initial conditions for the two assumed loop lengths in this paper, 50 Mm at left and 100 Mm at right.  The top plots show the electron and hydrogen density profiles, while the bottom plots show the initial temperature profile (hydrogen and electron temperatures are initially assumed to be equal).  At the apices of the loops, the initial radiative time-scales are about 10 and 20 minutes, respectively, while the initial conductive time-scales are about 100 and 140 minutes. \label{fig:initial}}
\end{figure*}

In Figure \ref{fig:initial}, we show the initial conditions for the two loop lengths assumed in this work (50 and 100 Mm).  The top plots show the initial density profiles, where solid blue marks the hydrogen density and dashed red marks the electron density.  The bottom plots show the initial temperature profiles, where initially the electron and hydrogen temperatures are assumed to be equal.  Additionally, the initial velocity everywhere is assumed to be zero.  The chromospheric temperature profile uses the VAL C model \citep{vernazza1981}, while the rest of the profiles are determined by solving the hydrostatic equations across the loop.  The photospheric boundaries are closed.

The initial radiative time-scales can be estimated from the temperatures and densities.  The radiative time-scale can be written $\tau_{R} = \Big(\frac{2}{\gamma -1}\Big)\frac{k_{B}T^{1-\alpha}}{\chi n}$ \citep{cargill1994}, where $\gamma = 5/3$, $k_{B}$ is the Boltzmann constant, and $\alpha$ and $\chi$ are constants that approximate the radiative loss function ($\propto \chi T^{\alpha}$; \citealt{rosner1978}).  In the temperature range of 0.4 to 1.5 MK, $\chi = 1.9 \times 10^{-22}$ and $\alpha = 0$ \citep{klimchuk2008}, so that the radiative time-scale becomes $\tau_{R} \approx 2.18 \times 10^{5} \Big(\frac{T}{n}\Big)$.  At the apices of the loops, this is about 10 and 20 minutes for the 50 and 100 Mm cases, respectively.  The conductive time-scales similarly found are around 100 and 140 minutes.

\section{Results}
\label{sec:results}

\begin{figure*}
\centering
\begin{minipage}[b]{\linewidth}
\includegraphics[width=0.5\linewidth]{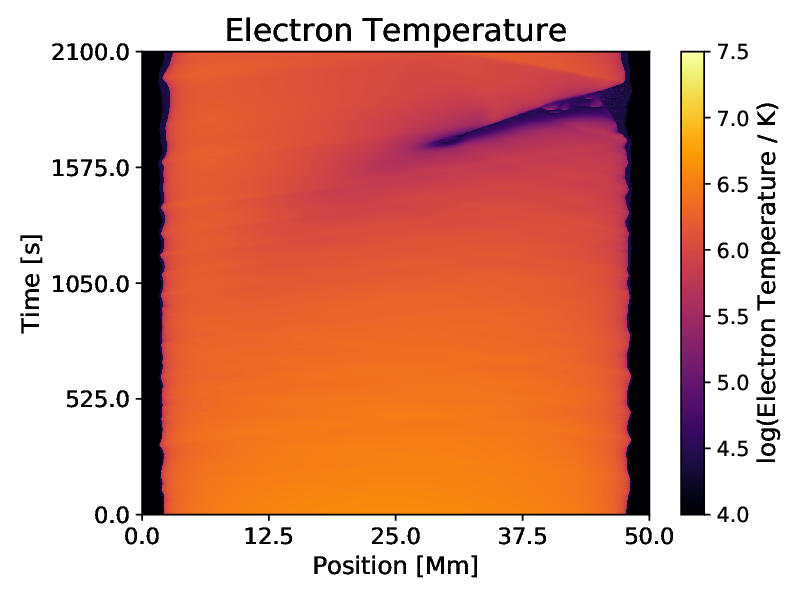}
\includegraphics[width=0.5\linewidth]{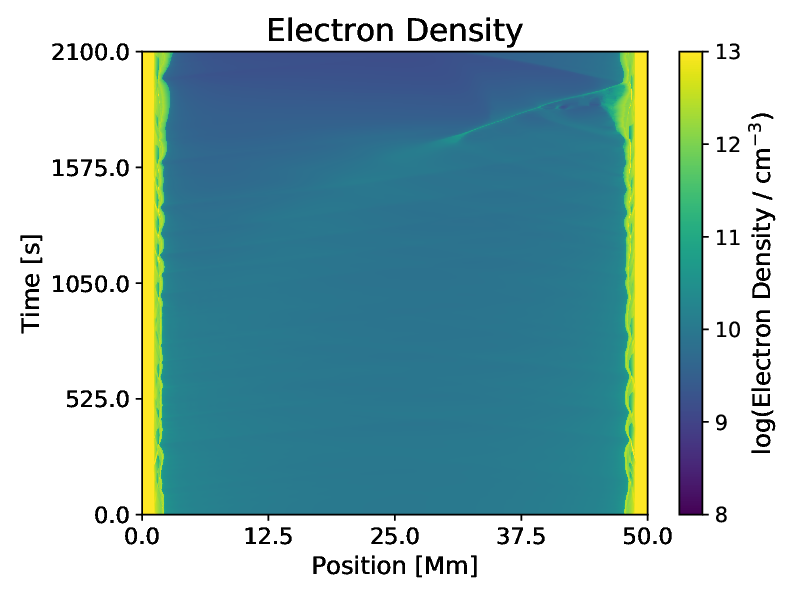}
\end{minipage}
\begin{minipage}[b]{\linewidth}
\includegraphics[width=0.5\linewidth]{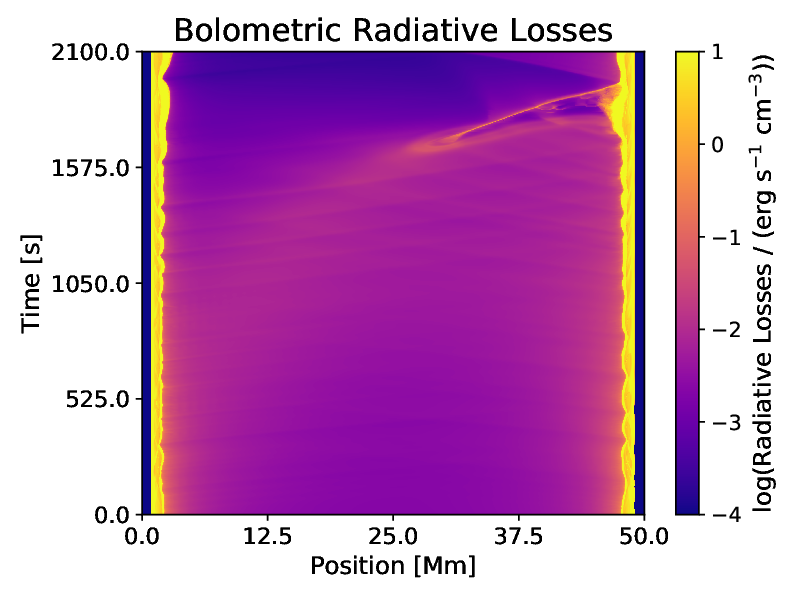}
\includegraphics[width=0.5\linewidth]{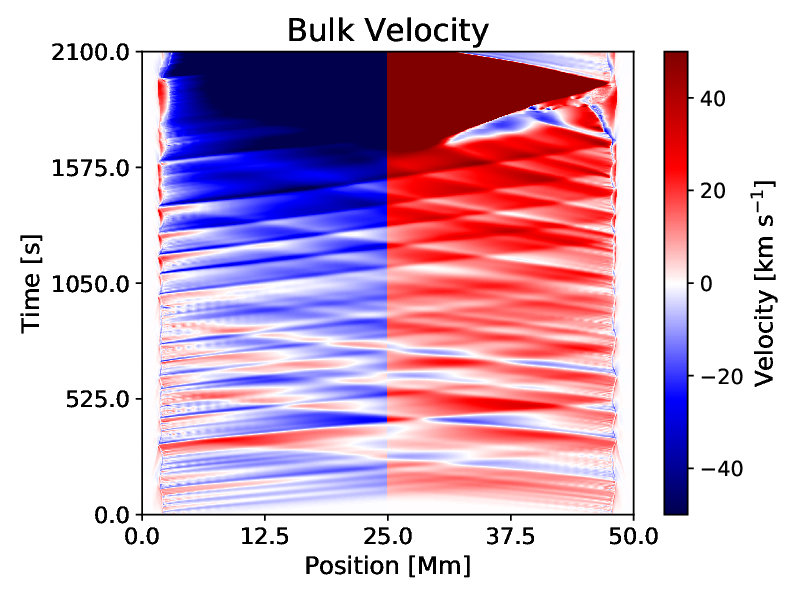}
\end{minipage}
\begin{minipage}[b]{\linewidth}
\includegraphics[width=0.5\linewidth]{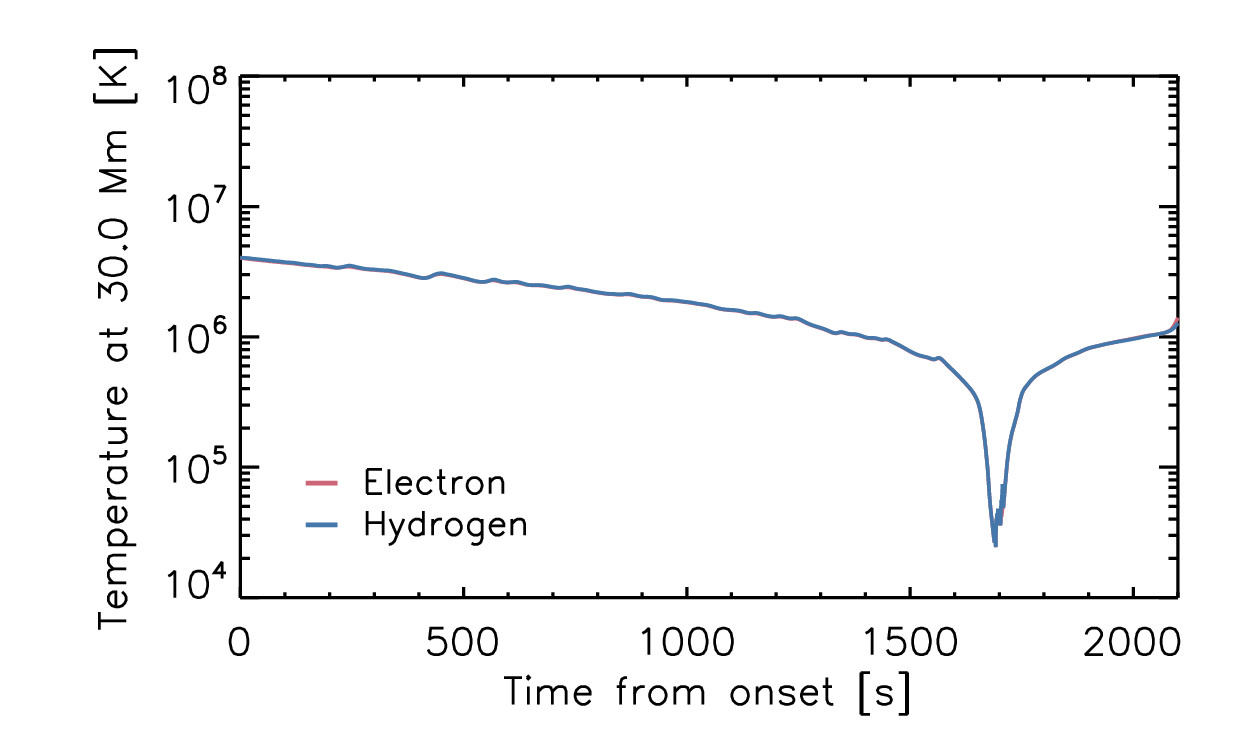}
\includegraphics[width=0.5\linewidth]{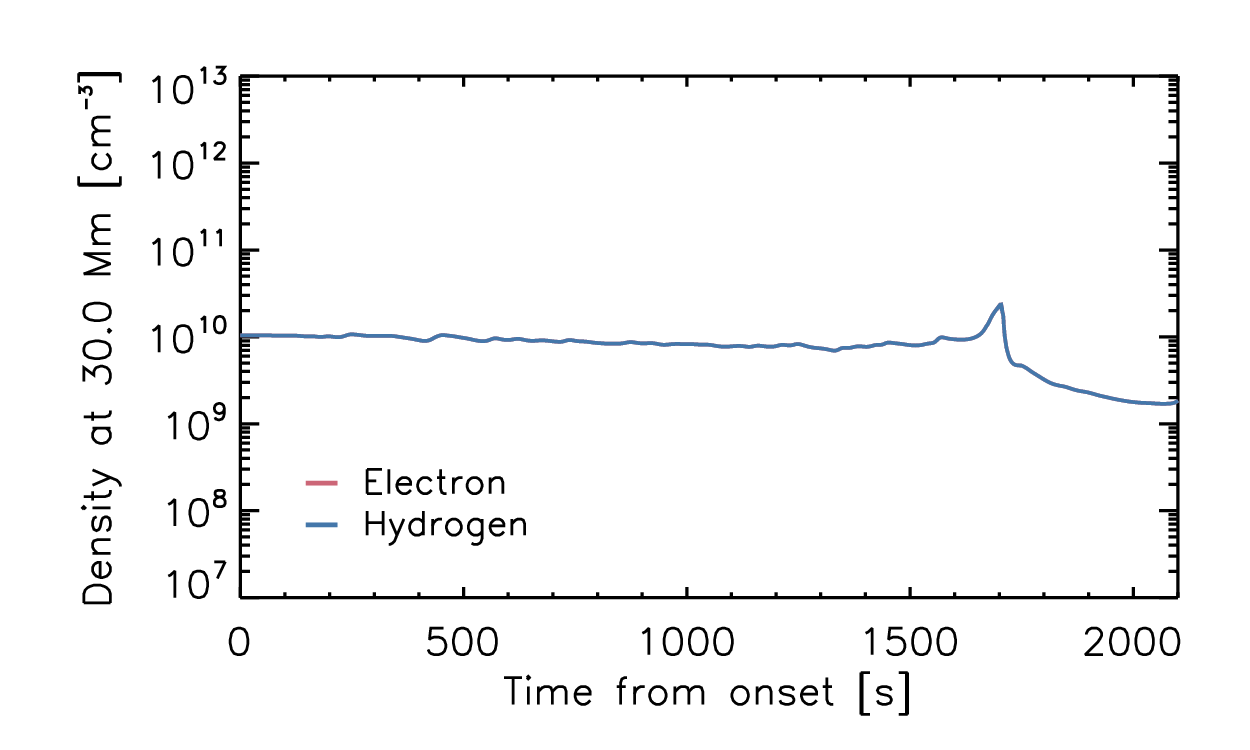}
\end{minipage}
\caption{An example of a coronal rain event in a standard (non-flaring) TNE scenario.  The plots show the evolution of a hot and dense loop subjected to asymmetric steady footpoint heating at the left-handed footpoint.  The x-axis of each plot shows the coordinate along the loop, the y-axis shows the change with time, while the color scales show the magnitude of each quantity.  From top left, the electron temperature, electron density, bolometric radiative losses, the bulk flow velocity (where red defines a flow away from the apex), and finally the evolution of the temperature and density at the position 30 Mm, approximately where the condensation begins to form.  The thermal instability develops after around 30 minutes of heating, causing a localized runaway cooling, forming a blob of high-density and low-temperature plasma that precipitates down the loop towards the right-handed footpoint.}
\label{fig:example_rain}
\end{figure*}

First, to illustrate coronal rain, we show a case in the standard TNE scenario, where a steady footpoint heating causes the formation of a rain event.  Figure \ref{fig:example_rain} shows the evolution of a 50\,Mm coronal loop subjected to a steady, asymmetric footpoint heating.  The heat is deposited entirely at the left-handed footpoint, centered at a height of 2 Mm above the photosphere, with a heating scale height of 3 Mm and volumetric heating rate of 0.05\,erg\,s$^{-1}$\,cm$^{-3}$, which heats the plasma to a temperature of 4 MK and eventually causes the formation of a thermal instability and localized runaway cooling.  Both symmetric and asymmetric heating can cause TNE and runaway cooling \citep{mikic2013,froment2018}, but we choose asymmetric heating in this case simply to ensure that the rain is likely to precipitate, rather than forming a stationary prominence (e.g. \citealt{karpen2005}).  A blob of high-density and low-temperature plasma forms in the corona, that then precipitates down towards the right-handed footpoint.  The plots, respectively, show the evolution of the electron temperature, electron density, bolometric radiative losses, and bulk flow velocity (red is defined as motion away from the apex) and the evolution of the temperature and density at the location 30 Mm, which is approximately where the condensation forms.  The x-axis shows the position coordinate along the loop, from footpoint to footpoint, where the apex is located at 25 Mm.  The y-axis similarly shows the change with time, while the color scale of each plot shows the magnitude of the given quantity.  Although the conduction is efficient at smoothing temperature gradients, the slow but persistent evaporative up-flows at the left-handed footpoint slowly build into an increased radiative loss rate ($\propto n^{2}$).  As the radiative losses grow from this density increase, the temperature decreases, which further increases the radiative loss rate.  This drives the runaway cooling necessary to form a condensation.  At around 1800 s (30 minutes), the thermal instability develops slightly rightward of the apex, forming at densities exceeding $10^{11}$\,cm$^{-3}$, which quickly cools to temperatures below $10^{5}$\,K.  This blob of plasma then falls towards the footpoint, where its impact with the chromosphere causes a small recoil of mass back into the corona.  

All of the properties of this event, including the morphology and dynamics of the rain, are similar to the quiescent coronal rain type \citep{antolin2020}.  In general, the dynamics and formation of condensations from TNE depend on the heating strength, duration, location, scale height, and asymmetry \citep{antiochos1991,muller2003,muller2004,muller2005,mikic2013,froment2018,johnston2019}.  The location and timing where condensations form \citep{karpen2005} as well as the rate at which they fall \citep{muller2005} depends on these heating properties, as well as properties of the magnetic geometry \citep{antiochos1980,karpen2001}.  

We now turn to electron beam heating.  In order to examine the capability of beams to produce TNE and/or coronal rain, we have run a large set of numerical experiments.  We have varied the properties of the electron beam: its energy flux ($10^{9}, 10^{10}, 10^{11}$\,erg\,s$^{-1}$\,cm$^{-2}$), low-energy cut-off (10, 20, 30, 50, 100\,keV), duration (10, 100\,s, and steady), and left-right asymmetry (0, 25, 50, 75, 100\%).  We have also varied the loop length ($2L = 50, 100$\,Mm) and cross-sectional area expansion (ratio of 1, 5, and 10 from footpoint to apex).  We assume in all the simulations in this work a spectral index $\delta = 5$ since it does not strongly impact the evolution of the loops.  We do not include a coronal background heating term at all. \citet{johnston2019} showed that if the background heating is large enough to compensate for the radiative losses, then the formation of coronal rain and TNE cycles is inhibited.

\begin{figure*}
\centering
\begin{minipage}[b]{\linewidth}
\includegraphics[width=0.5\linewidth]{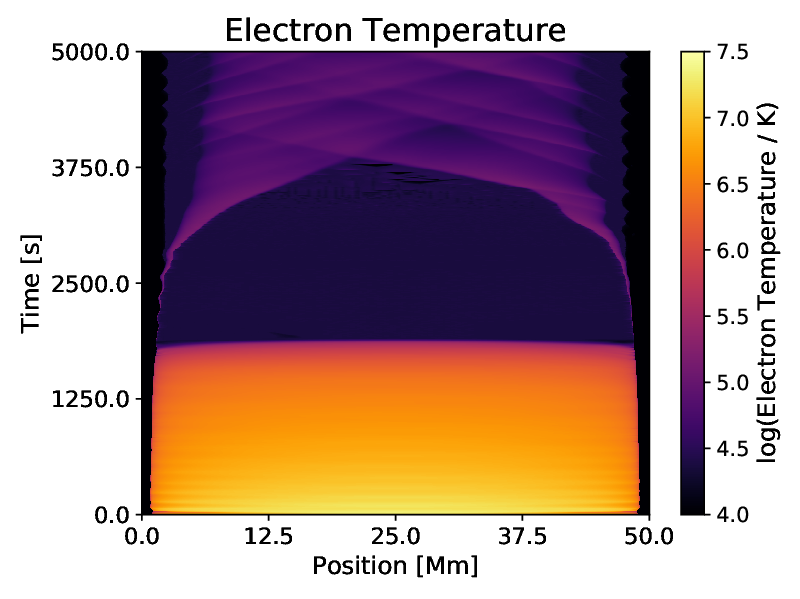}
\includegraphics[width=0.5\linewidth]{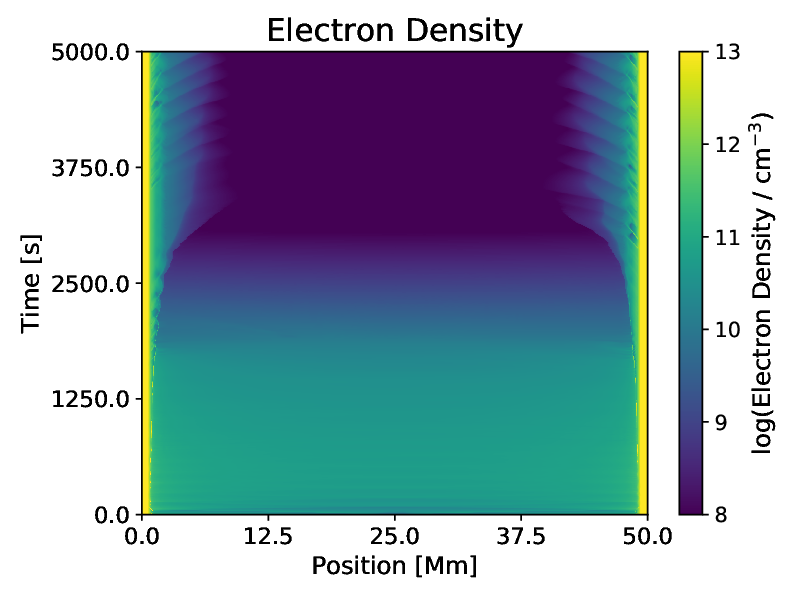}
\end{minipage}
\begin{minipage}[b]{\linewidth}
\includegraphics[width=0.5\linewidth]{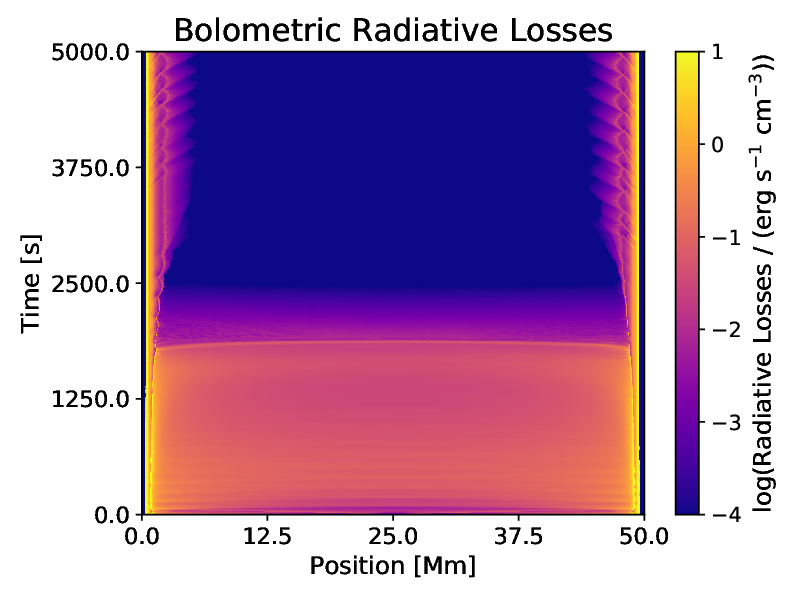}
\includegraphics[width=0.5\linewidth]{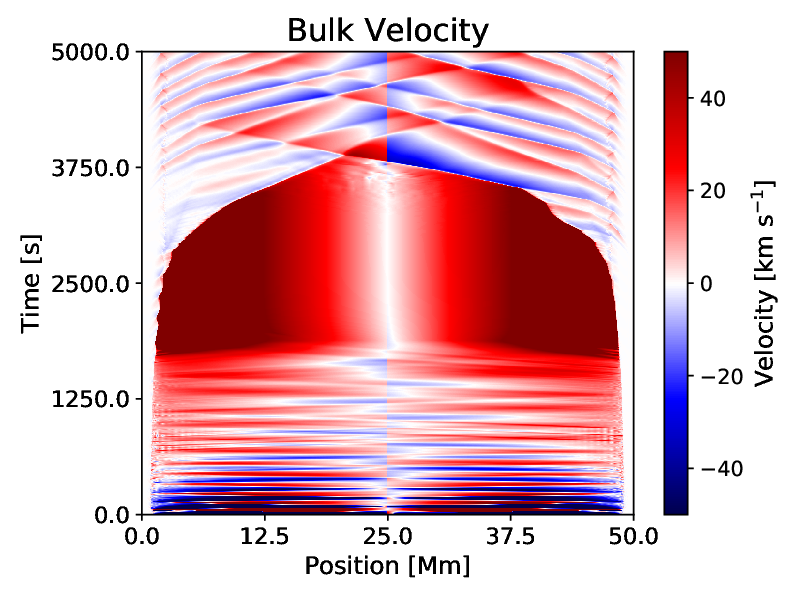}
\end{minipage}
\begin{minipage}[b]{\linewidth}
\includegraphics[width=0.5\linewidth]{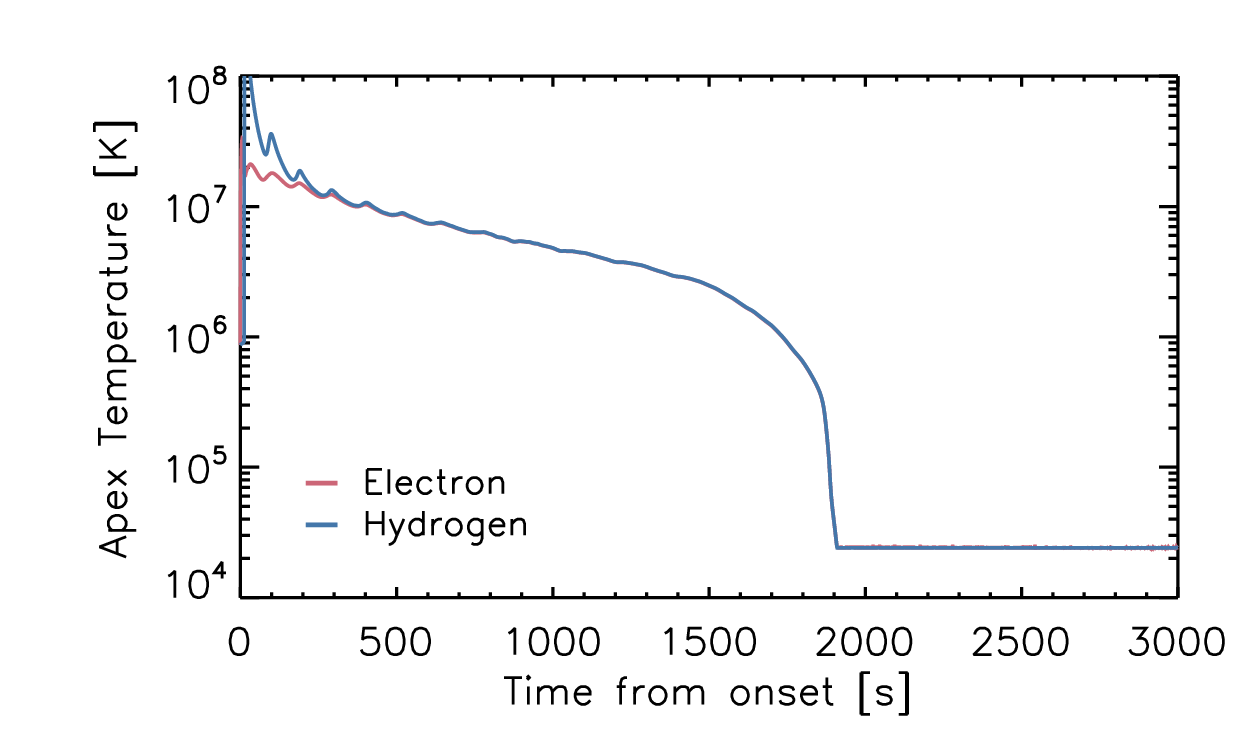}
\includegraphics[width=0.5\linewidth]{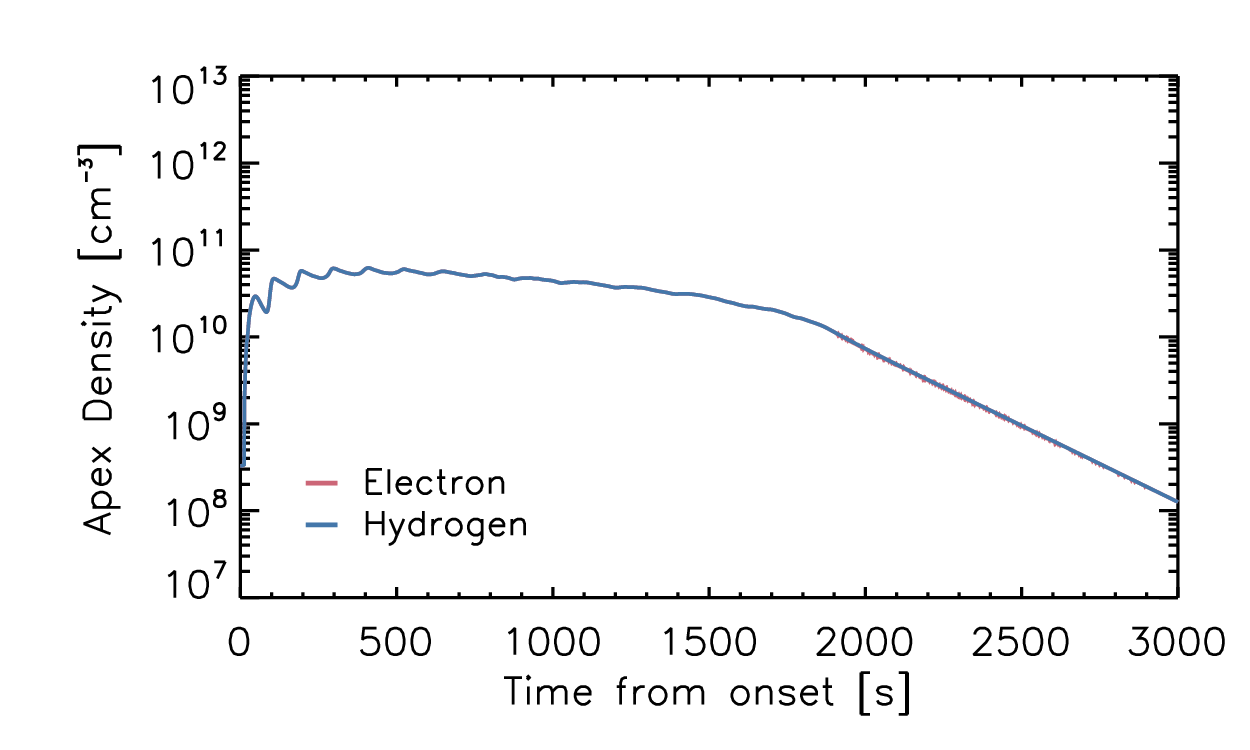}
\end{minipage}
\caption{An example of a flaring loop heated strongly by a short, powerful, and symmetric electron beam, using parameters often considered typical in flare simulations.  The loop quickly heats up, fills with plasma, and in less than 30 minutes begins to cool and drain.  No coronal rain forms.  Respectively, from top left, the electron temperature, electron density, bolometric radiative losses, and bulk flow velocity (where red defines a flow away from the apex), and the values of the apex temperatures and densities as a function of time.}
\label{fig:example_flare}
\end{figure*}

We begin with a detailed example of a flare simulation, using parameters typical of those assumed in modeling papers.  Consider the case of a relatively short loop (50 Mm) with uniform cross section, heated strongly by a symmetric electron beam for 10\,s, with low-energy cut-off $10$\,keV and energy flux of $10^{11}$\,erg\,s$^{-1}$\,cm$^{-2}$.  Figure \ref{fig:example_flare} shows the evolution of the (respectively from top left) electron temperature, electron density, bolometric radiative losses, the bulk flow velocity, and the evolution of the temperatures and densities at the apex of the loop.  The velocity plot shows a flow towards the apex as blue, away from the apex as red.  The temperature of the loop quickly rises to above 20\,MK at its maximum, which slowly cools following the heating period.  The initial burst of energy drives a strong evaporative flow into the corona, raising the density to around 10$^{11}$\,cm$^{-3}$, strongly increasing the radiative loss rate.  Eventually, there is an onset of a global catastrophic cooling \citep{cargill2013}, where the loop \textit{as a whole} drops to temperatures below $10^{5}$\,K, while the density remains roughly constant, which then drives a strong outflow of material from the loop.  Because there is no background heating assumed in the corona, the loop does not fill or heat again after this collapse.  Pressure waves form and bounce across the corona during the cooling phase, but these waves flow up the loops as well as down, which is not consistent with observed coronal rain.  Additionally, the pressure waves are not distinctly lower in temperature than the rest of the loop, and so would appear as brightenings at the same wavelengths (with a sufficiently sensitive instrument).  Before the collapse of the loop, because there are no localized condensations of cool material, we see that this example fails to produce rain.

In order to demonstrate that the lack of coronal rain is not due to the choice of parameters, we now explore the parameter space, varying both beam and loop parameters.  We find no case with coronal rain.

\begin{figure*}
\centering
\begin{minipage}[b]{\linewidth}
\includegraphics[width=0.33\linewidth]{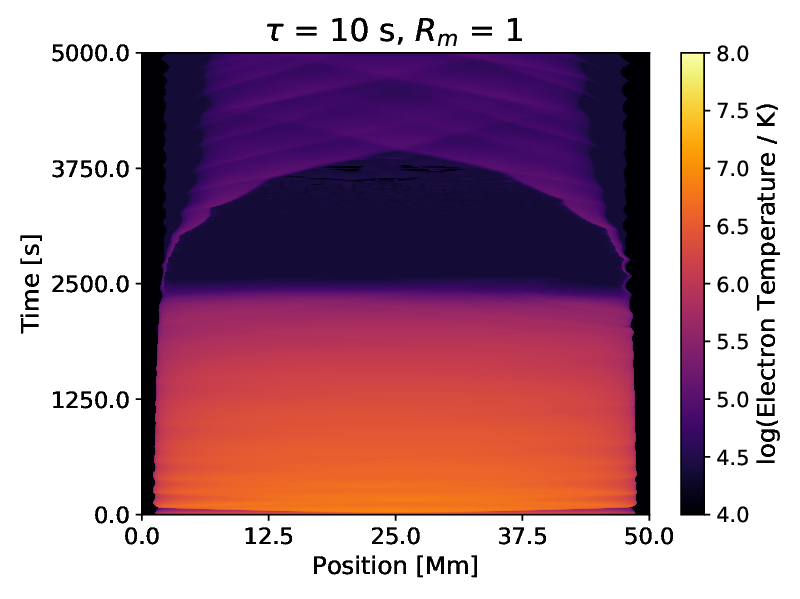}
\includegraphics[width=0.33\linewidth]{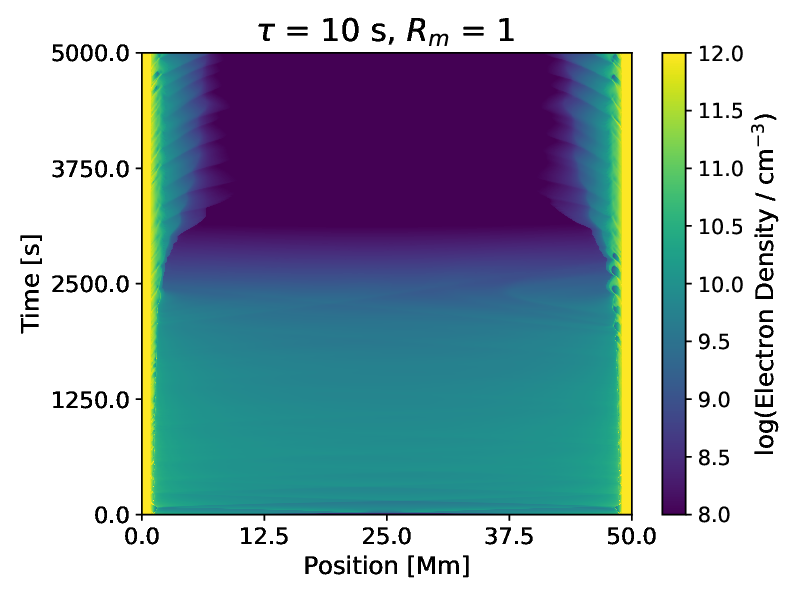}
\includegraphics[width=0.33\linewidth]{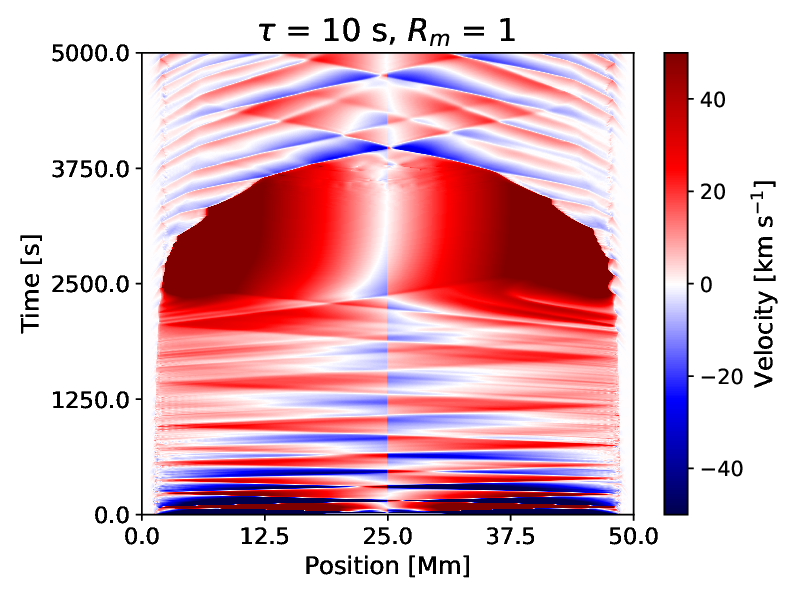}
\end{minipage}
\begin{minipage}[b]{\linewidth}
\includegraphics[width=0.33\linewidth]{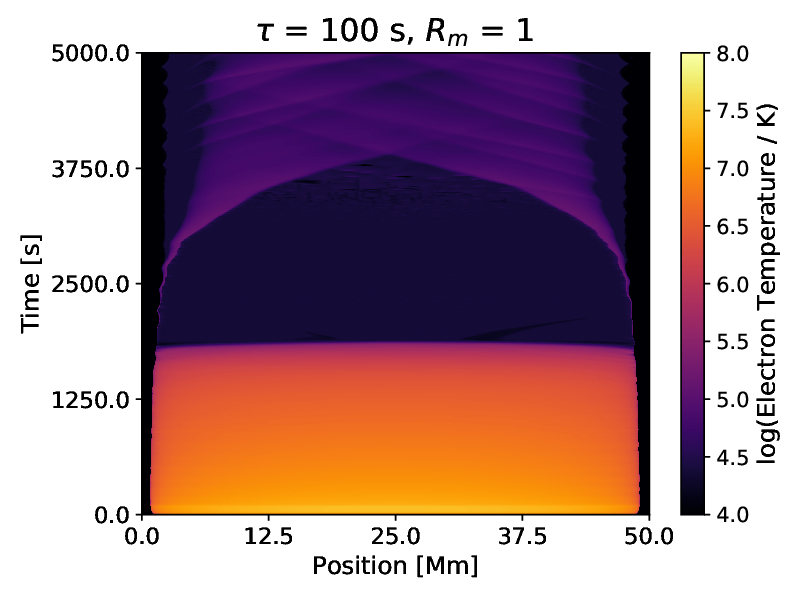}
\includegraphics[width=0.33\linewidth]{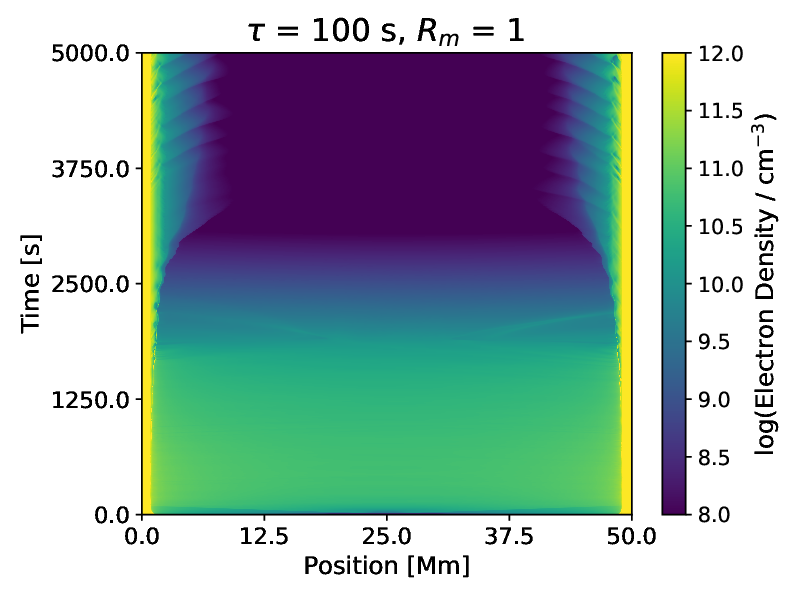}
\includegraphics[width=0.33\linewidth]{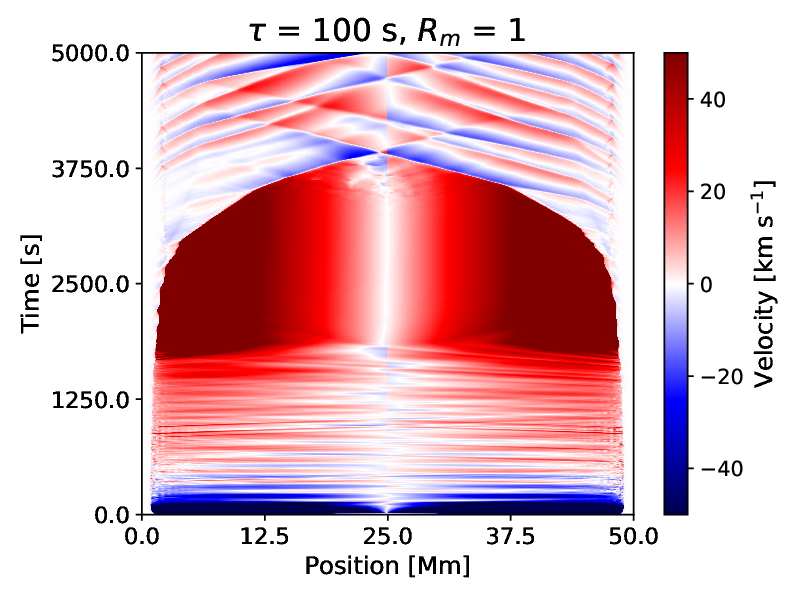}
\end{minipage}
\begin{minipage}[b]{\linewidth}
\includegraphics[width=0.33\linewidth]{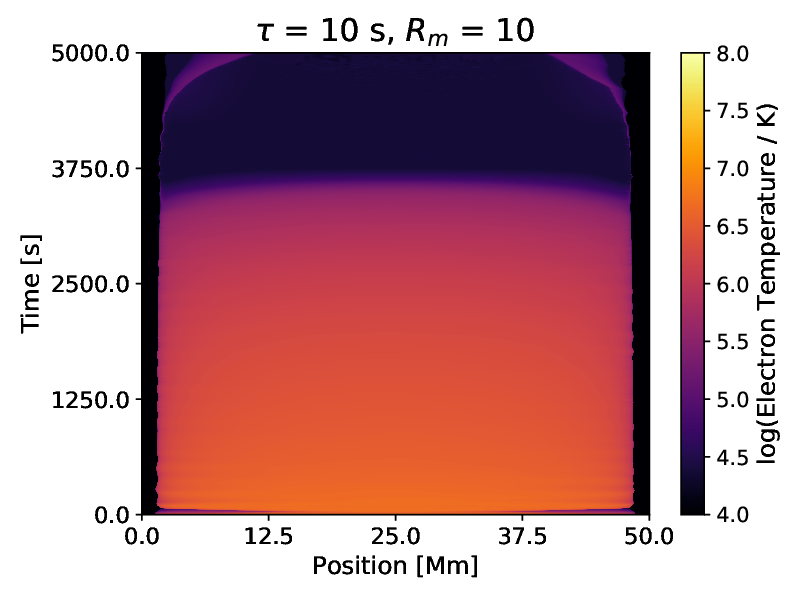}
\includegraphics[width=0.33\linewidth]{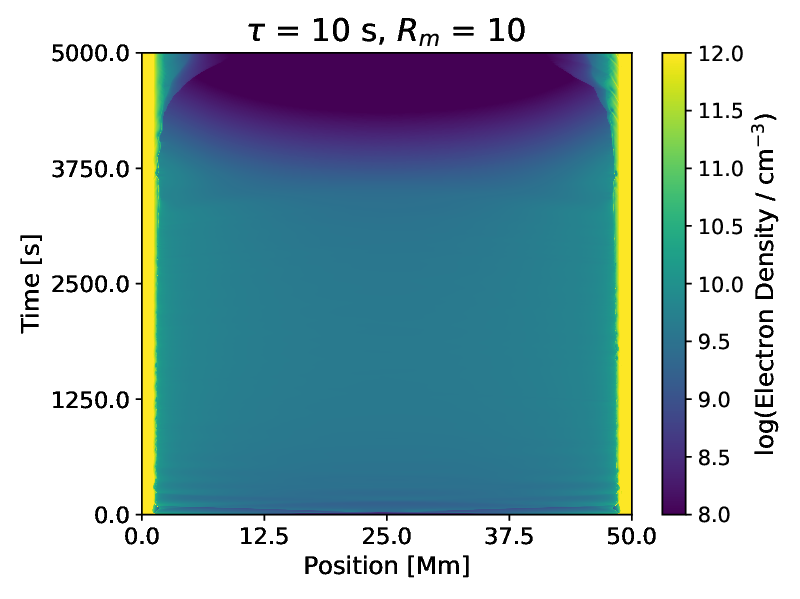}
\includegraphics[width=0.33\linewidth]{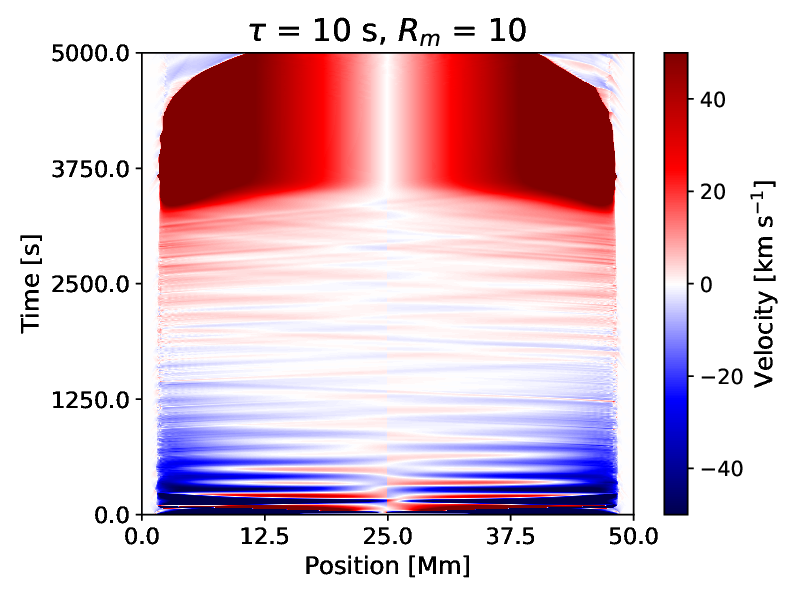}
\end{minipage}
\begin{minipage}[b]{\linewidth}
\includegraphics[width=0.33\linewidth]{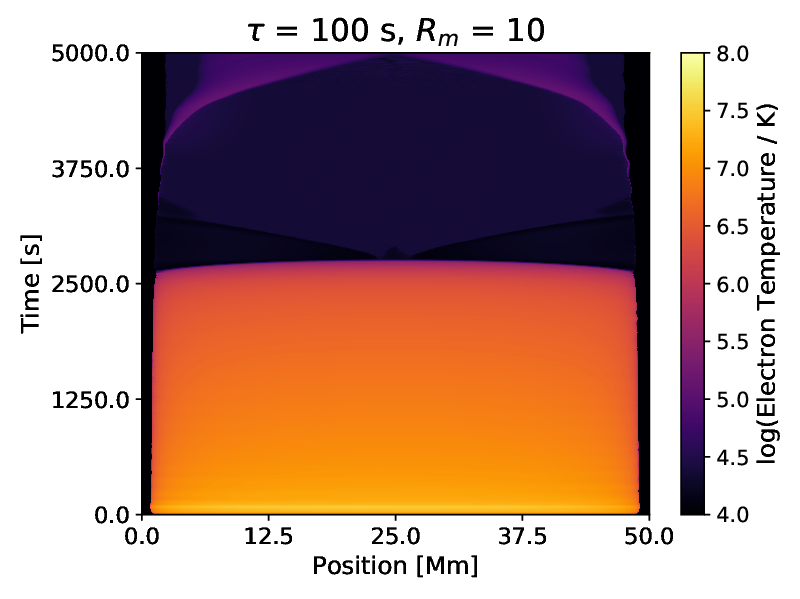}
\includegraphics[width=0.33\linewidth]{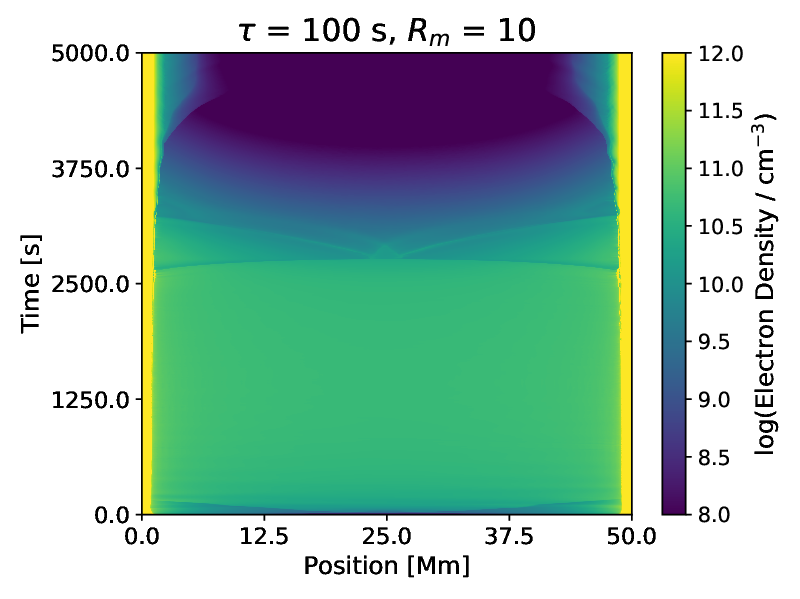}
\includegraphics[width=0.33\linewidth]{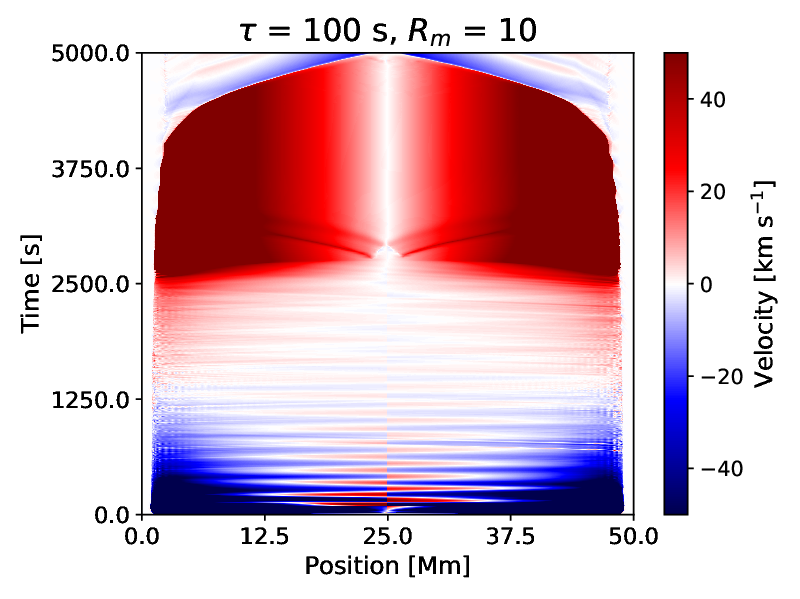}
\end{minipage}
\caption{Symmetric beam heating with durations of 10 and 100\,s, and magnetic mirror ratios $R_{m} = $1 (uniform) and 10, as labeled.  Each column respectively shows the electron temperature, electron density, and bulk flow velocity along the loops as a function of time.  No coronal rain forms in any of these cases.}
\label{fig:durations}
\end{figure*}

First, consider the effect of the beam duration.  In Figure \ref{fig:durations}, we show the evolution of 50\,Mm loops heated with symmetric beams, with durations of 10 and 100\,s, with magnetic mirror ratio (expansion of cross-sectional area from footpoint to apex) $R_{m} = $1 (uniform) and 10.  The columns show the electron temperature, electron density, and bulk flow velocity, respectively.  These simulations assume a constant energy flux of $F_{0} = 10^{10}$\,erg\,s$^{-1}$\,cm$^{-2}$ and low-energy cut-off $E_{c} = 10$\,keV.  The longer heating duration (equivalently, higher energy input) causes the loop to reach higher temperatures and densities, which in turn causes it to cool faster.  The effect of an expanding cross-section effectively reduces thermal conduction and thereby causes energy to be retained in the corona for longer, slowing the overall cooling of the loop.  As in Figure \ref{fig:example_flare}, there are pressure waves that form, but these are not cooler than the surrounding loop, and so are not consistent with coronal rain.  In none of these cases, therefore, does a coronal condensation form that could be consistent with rain.  

\begin{figure*}
\centering
\begin{minipage}[b]{\linewidth}
\includegraphics[width=0.33\linewidth]{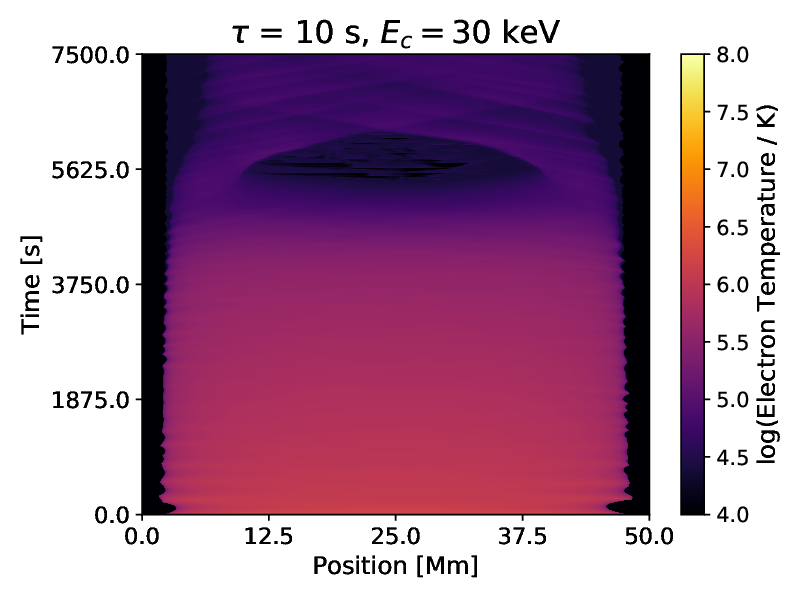}
\includegraphics[width=0.33\linewidth]{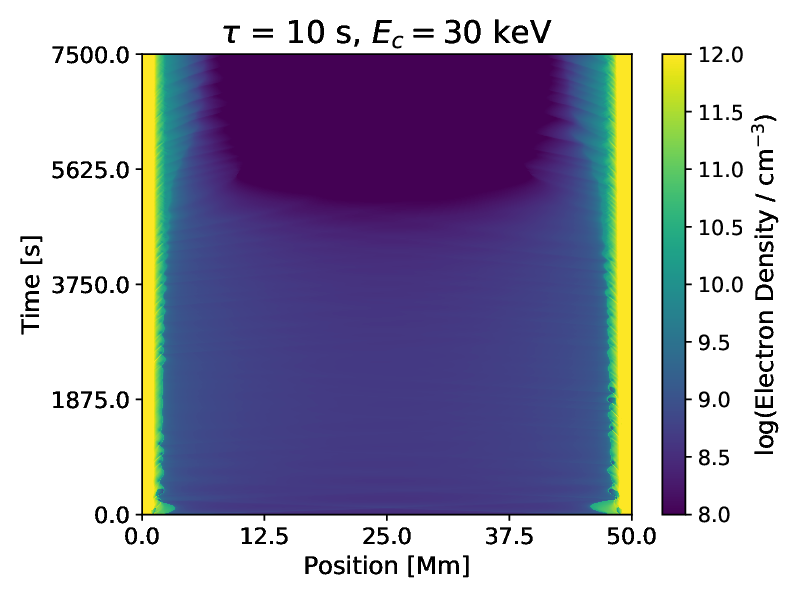}
\includegraphics[width=0.33\linewidth]{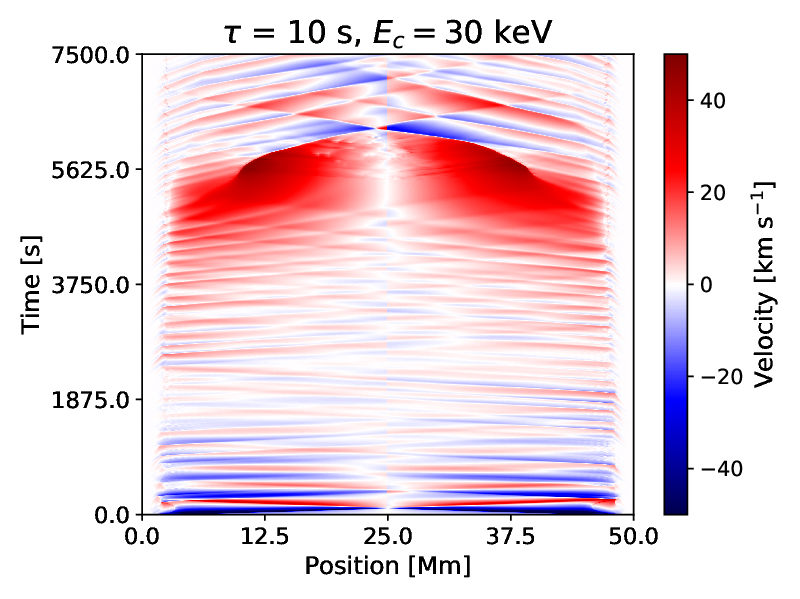}
\end{minipage}
\begin{minipage}[b]{\linewidth}
\includegraphics[width=0.33\linewidth]{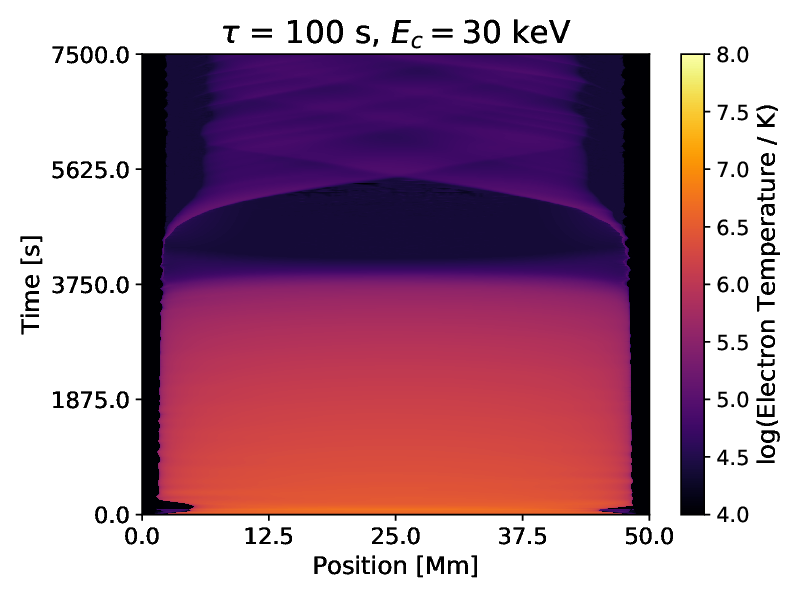}
\includegraphics[width=0.33\linewidth]{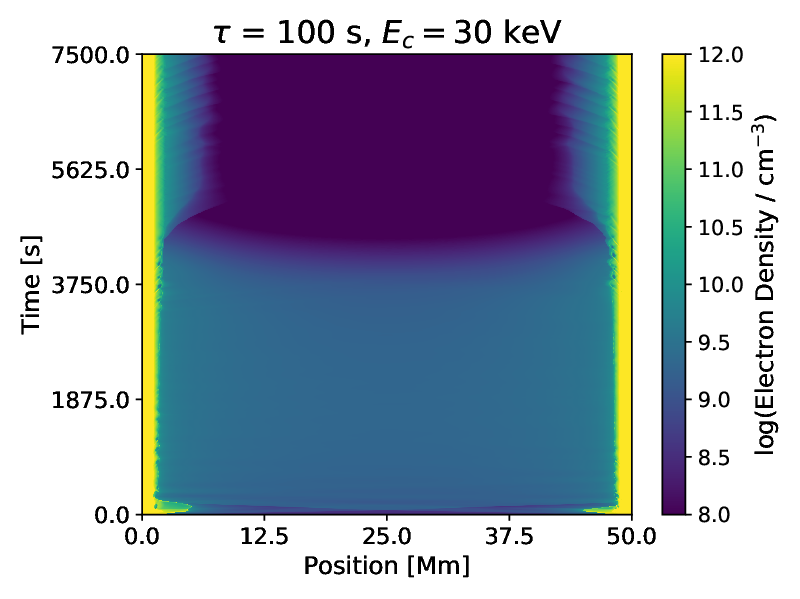}
\includegraphics[width=0.33\linewidth]{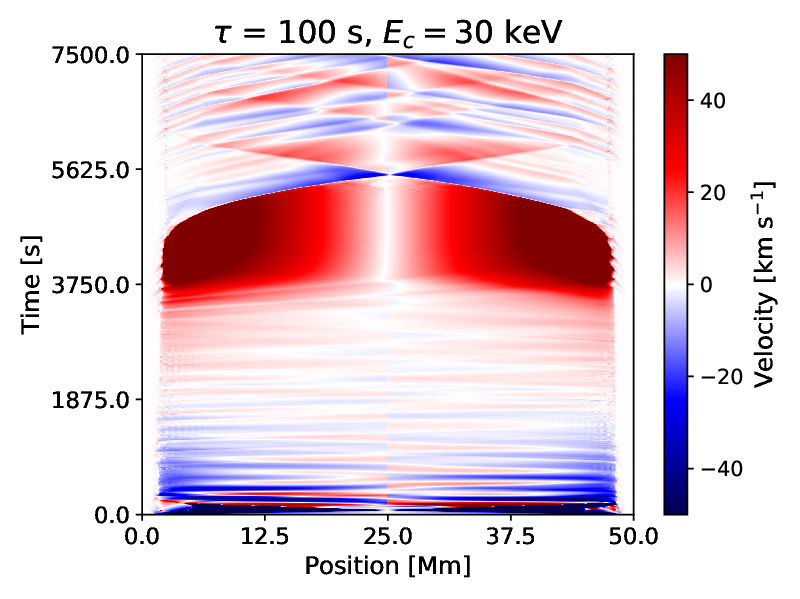}
\end{minipage}
\begin{minipage}[b]{\linewidth}
\includegraphics[width=0.33\linewidth]{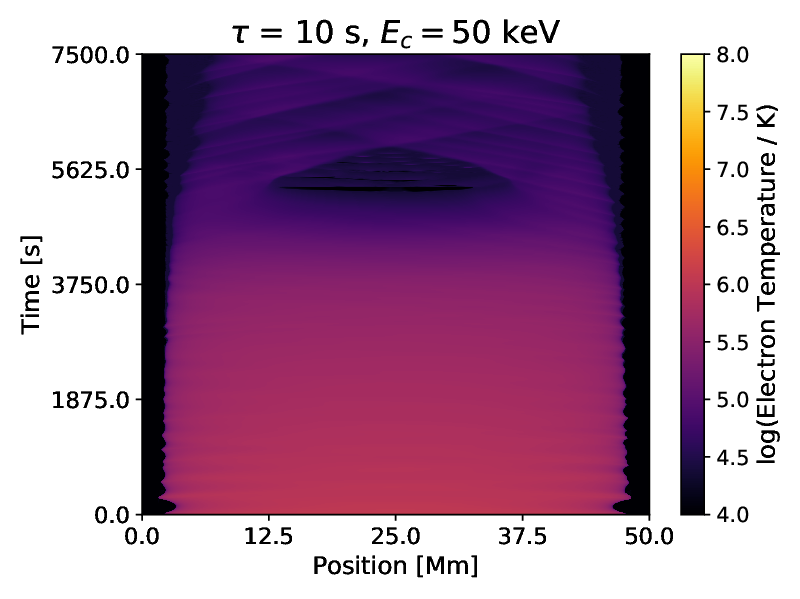}
\includegraphics[width=0.33\linewidth]{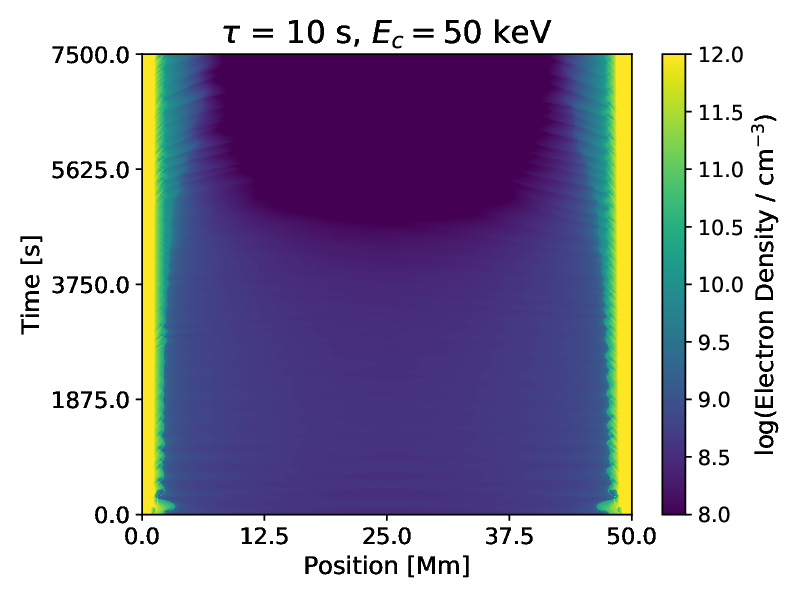}
\includegraphics[width=0.33\linewidth]{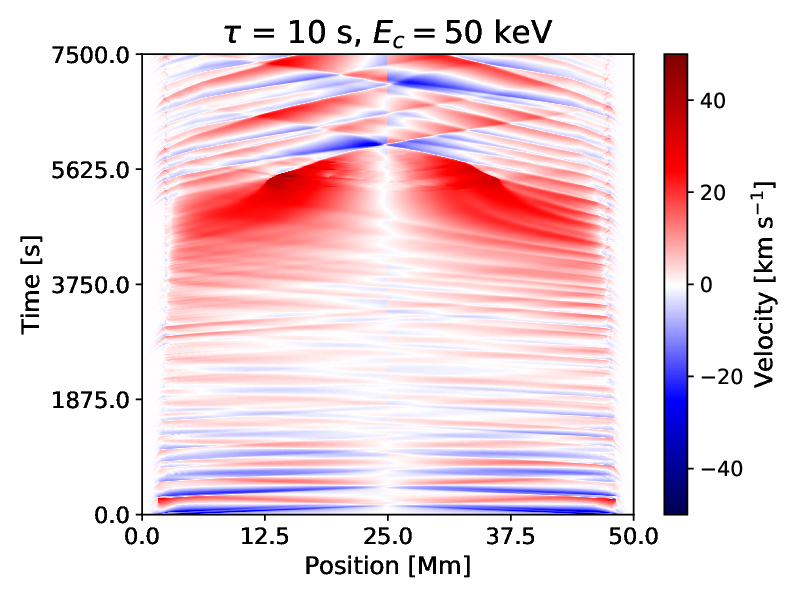}
\end{minipage}
\begin{minipage}[b]{\linewidth}
\includegraphics[width=0.33\linewidth]{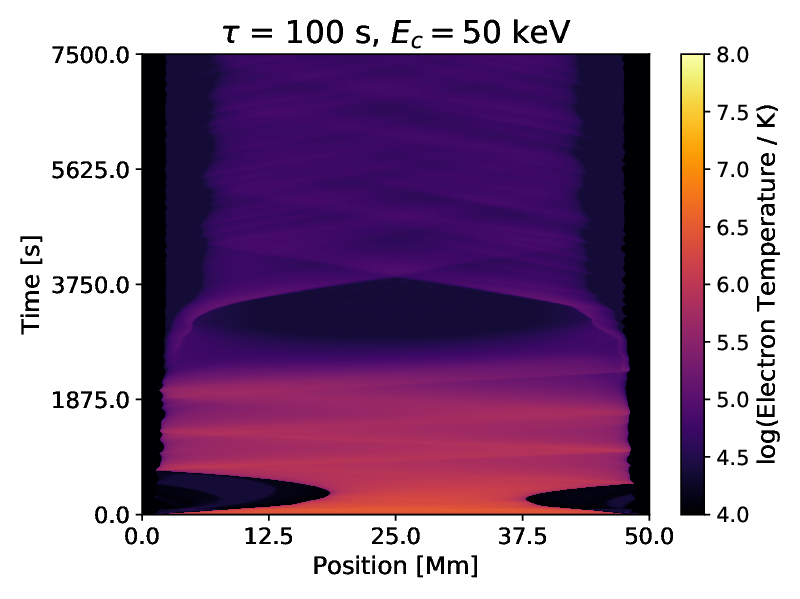}
\includegraphics[width=0.33\linewidth]{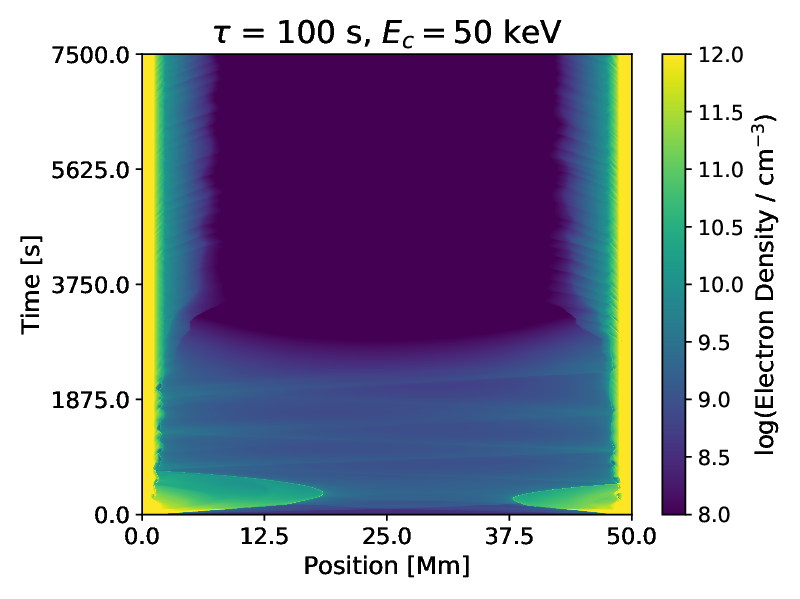}
\includegraphics[width=0.33\linewidth]{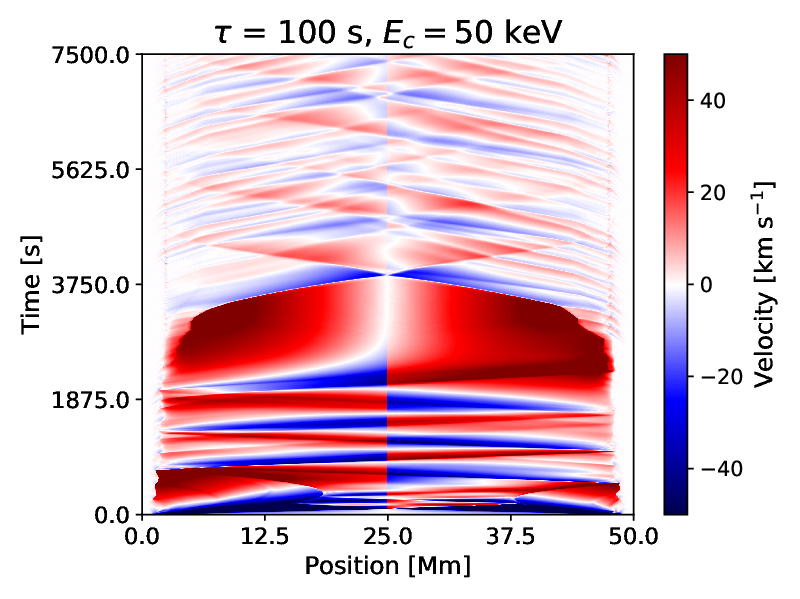}
\end{minipage}
\caption{Similar to Figure \ref{fig:durations}, but the low-energy cut-off $E_{c} = $ 30 and 50\,keV, heating durations 10 and 100\,s, and fixed mirror ratio $R_{m} = 10$.  Once again, no rain forms.}
\label{fig:cutoffs}
\end{figure*}

The precise depth of heating has also been shown to play a significant role in the formation of TNE \citep{froment2018}, so we now focus on the role of the low-energy cut-off $E_{c}$.  In Figure \ref{fig:cutoffs}, we show simulations with varying cut-off $E_{c} = $[20, 50]\,keV, for heating durations of 10 and 100\,s.  As before, we assume a fixed energy flux of $10^{10}$\,erg\,s$^{-1}$\,cm$^{-2}$.  We use a large magnetic mirror ratio $R_{m} = 10$.  Because the cut-off is higher, the energy carried by the beam is deposited at a lower average depth in the chromosphere, resulting in less heating than the previous case.  The loops do not get as hot or dense, and take longer to cool as a result.  In the case of $E_{c} = 50$\,keV, there is even a surge of cool material that wells up from the chromosphere, looking like a chromospheric jet.  However, in none of these cases do we observe a coronal condensation or evidence of rain.   

\begin{figure*}
\centering
\begin{minipage}[b]{\linewidth}
\includegraphics[width=0.33\linewidth]{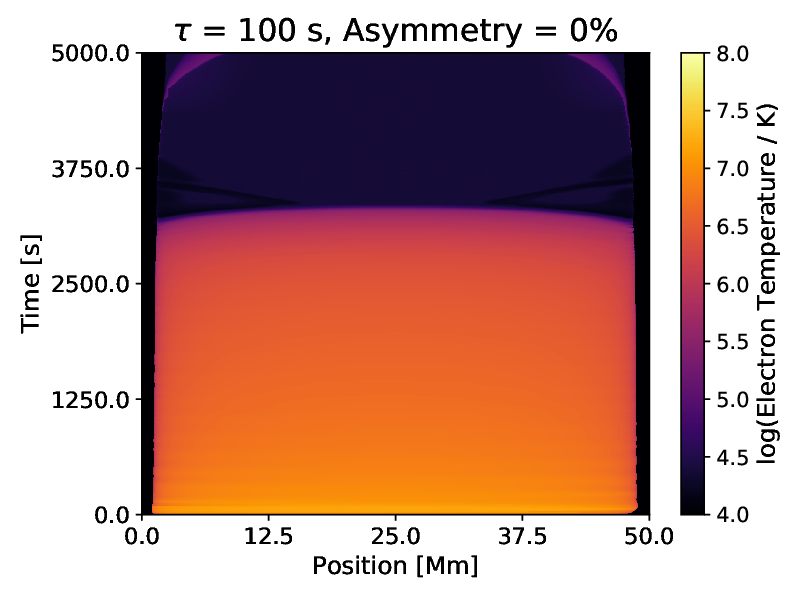}
\includegraphics[width=0.33\linewidth]{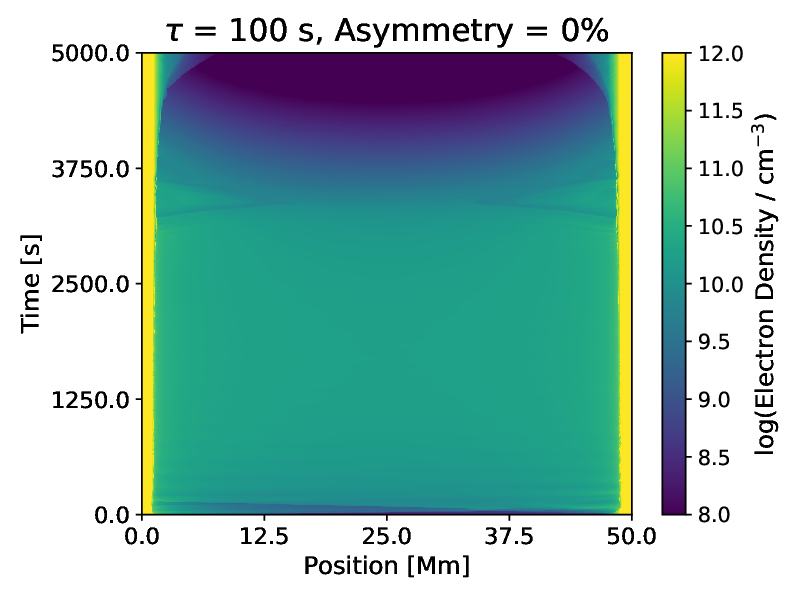}
\includegraphics[width=0.33\linewidth]{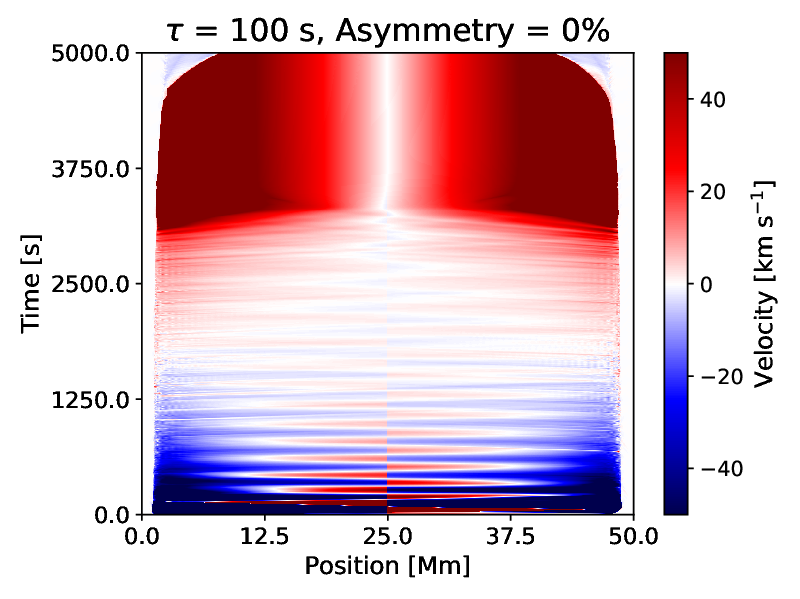}
\end{minipage}
\begin{minipage}[b]{\linewidth}
\includegraphics[width=0.33\linewidth]{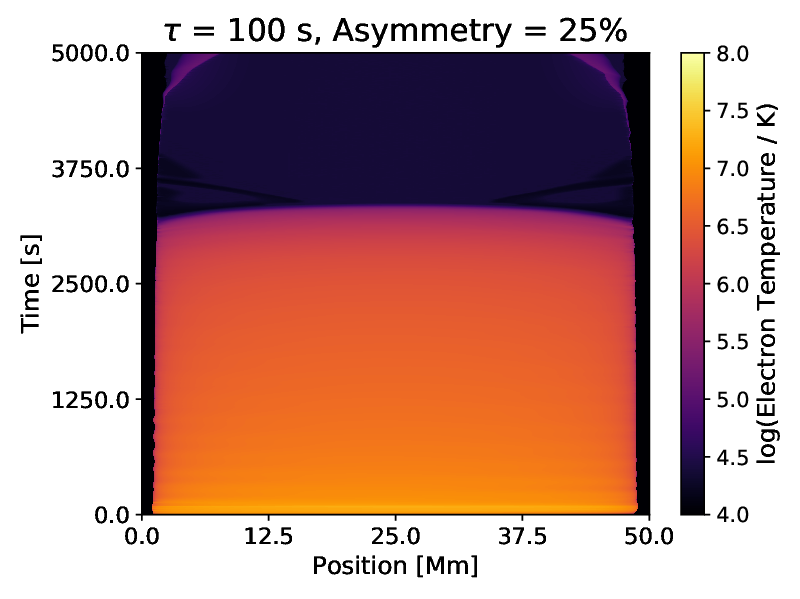}
\includegraphics[width=0.33\linewidth]{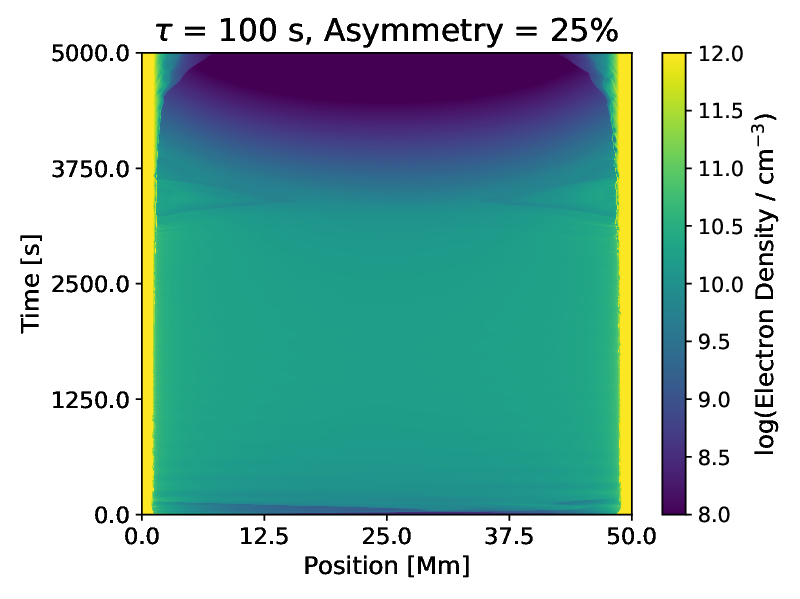}
\includegraphics[width=0.33\linewidth]{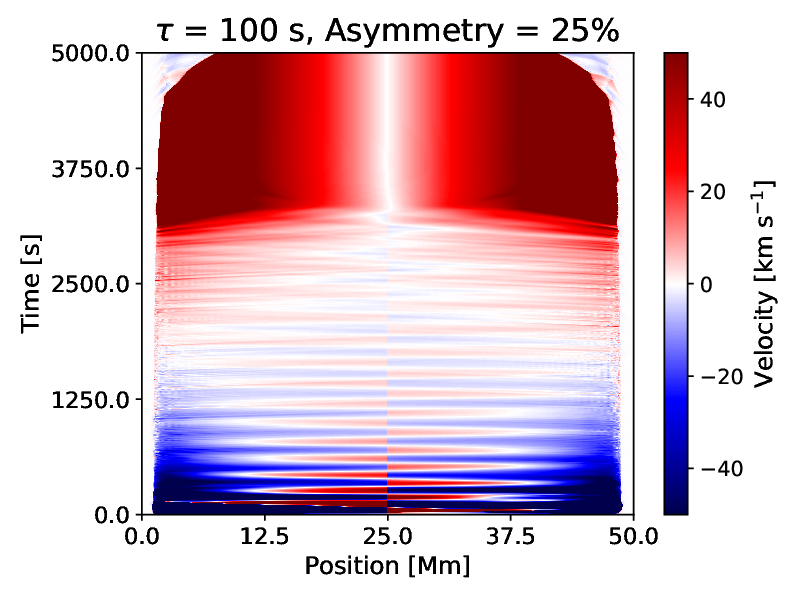}
\end{minipage}
\begin{minipage}[b]{\linewidth}
\includegraphics[width=0.33\linewidth]{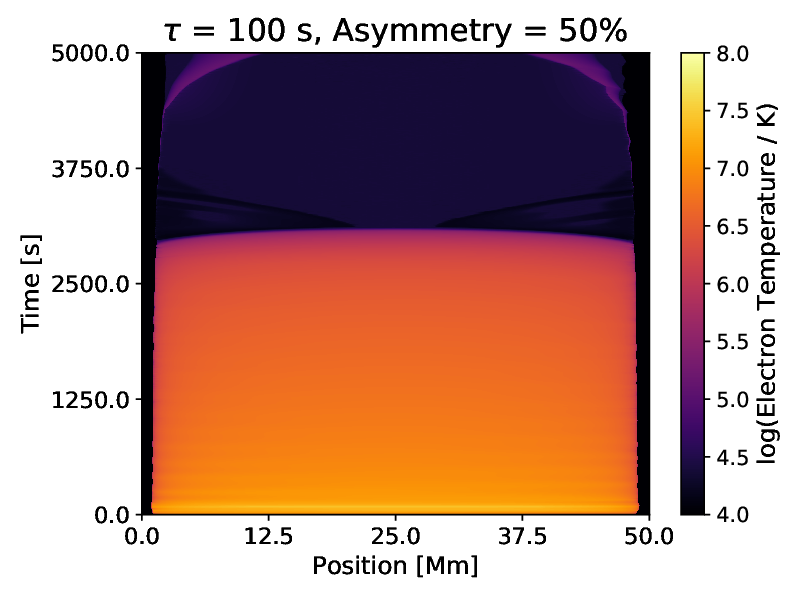}
\includegraphics[width=0.33\linewidth]{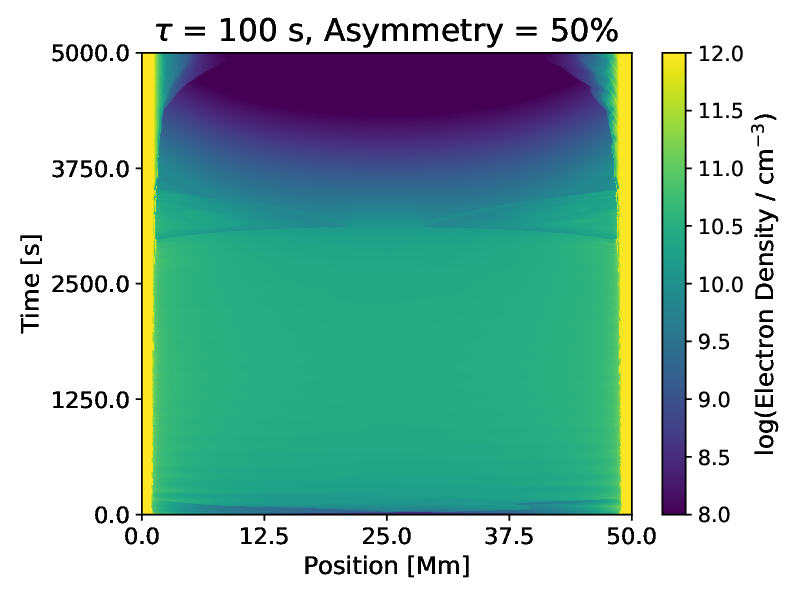}
\includegraphics[width=0.33\linewidth]{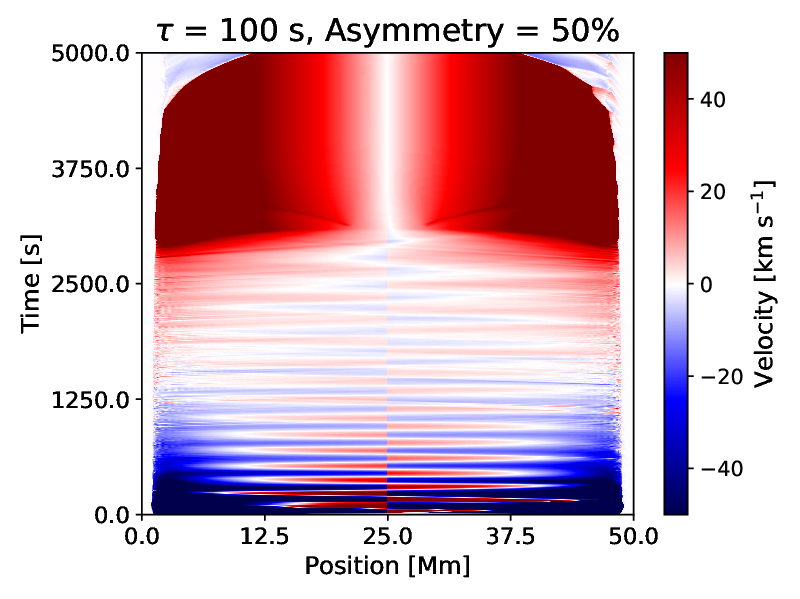}
\end{minipage}
\begin{minipage}[b]{\linewidth}
\includegraphics[width=0.33\linewidth]{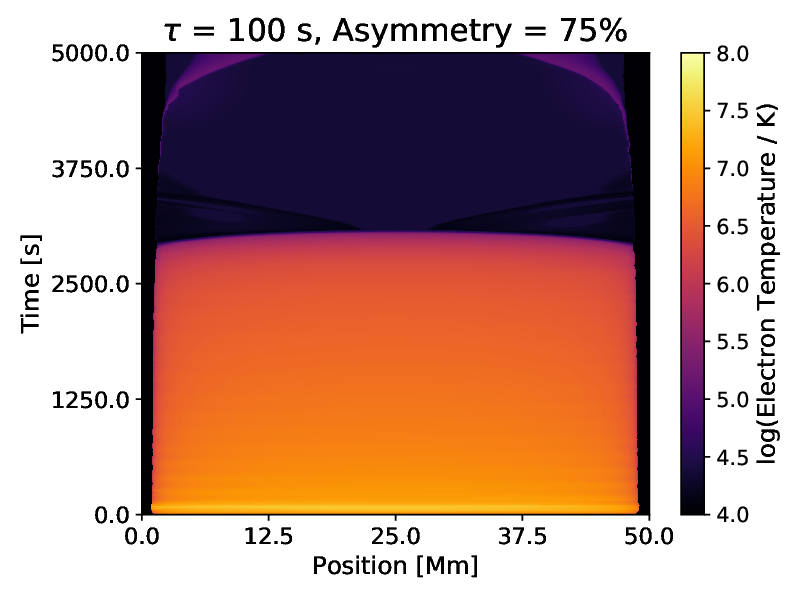}
\includegraphics[width=0.33\linewidth]{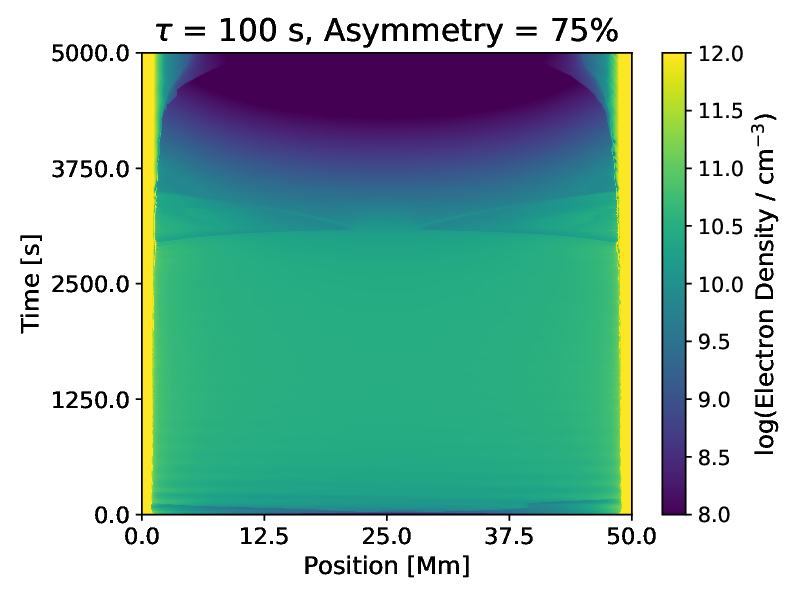}
\includegraphics[width=0.33\linewidth]{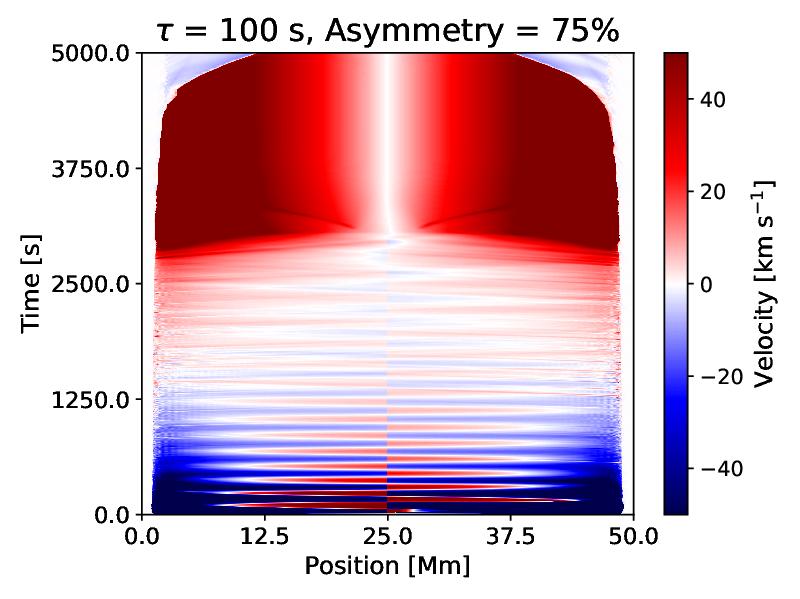}
\end{minipage}
\caption{Similar to Figure \ref{fig:durations}, but with asymmetric electron beams, where the right-hand leg receives 0, 25, 50, and 75\% as much energy as the left. }
\label{fig:asymmetric}
\end{figure*}

We next turn our attention to asymmetries in the heating, where one half of the loop receives more energy than the other half, which can drive siphon flows that may impact the probability of coronal rain \citep{klimchuk2019}.  In Figure \ref{fig:asymmetric}, we therefore show four cases with asymmetries of 0, 25, 50, and 75\% (\textit{i.e.} the right-hand leg of the loop receives x\% as much energy as the left).  We assume that the energy flux is $F_{0} = 10^{10}$\,erg\,s$^{-1}$\,cm$^{-2}$ and cut-off $E_{c} = 10$\,keV as before, with a heating duration of 100\,s, and magnetic mirror ratio $R_{m} = 10$.  While there are differences between the simulations, particularly early on during the heating period, the beams likely do not last long enough for large asymmetries to form in the loop once heating ceases.  Furthermore, the efficiency of thermal conduction efficiently smooths out any gradients in the temperature.  Initially, the conductive time-scale, $\tau_{c} = \Big(\frac{2}{\gamma-1}\Big)\frac{n k_{B} L^{2}}{\kappa_{0} T^{5/2}}$, is around 100 minutes in these loops.  After the onset of heating, the temperature rises to around 20 MK and the density to around $5 \times 10^{10}$ cm$^{-3}$, reducing the conductive time-scale to around 5 minutes.  The conduction therefore quickly acts to smooth out the variations in temperature that may have been present due to asymmetries in the heating.  The overall effect is that these loops evolve much like the previous cases, and no rain forms.

Rain does not form in any of these cases.  The full simulation set with many more permutations of parameters is available for the interested reader in the supplemental material, with a significantly larger number of simulations than presented here.  All of the beam and loop parameters were varied, but based on the reasoning in Section \ref{sec:beams}, we do not expect that there exists some combination of reasonable parameters that can produce rain with an electron beam, and that there must be an additional mechanism involved.

\section{Discussion}
\label{sec:discussion}

The numerical experiments have failed to find any evidence for the formation or occurrence of TNE or coronal rain in loops subjected to only heating by an electron beam.  We have run a large set of simulations designed to be comprehensive enough to check reasonable combinations of parameters to see if rain can be produced from electron beam heating.  In all cases, the loop collapses before any rain forms.  It is unlikely that electron beams on their own are capable of producing TNE for the two reasons stated earlier: (1) beams do not last long enough to produce the thermal instability, and (2) the location of energy deposition by electron beams does not remain fixed at the footpoints.  The question remains: how does coronal rain form in solar flares?  More generally, how does rain form in impulsive events?

\begin{figure*}
\centering
\begin{minipage}[b]{\linewidth}
\includegraphics[width=0.5\linewidth]{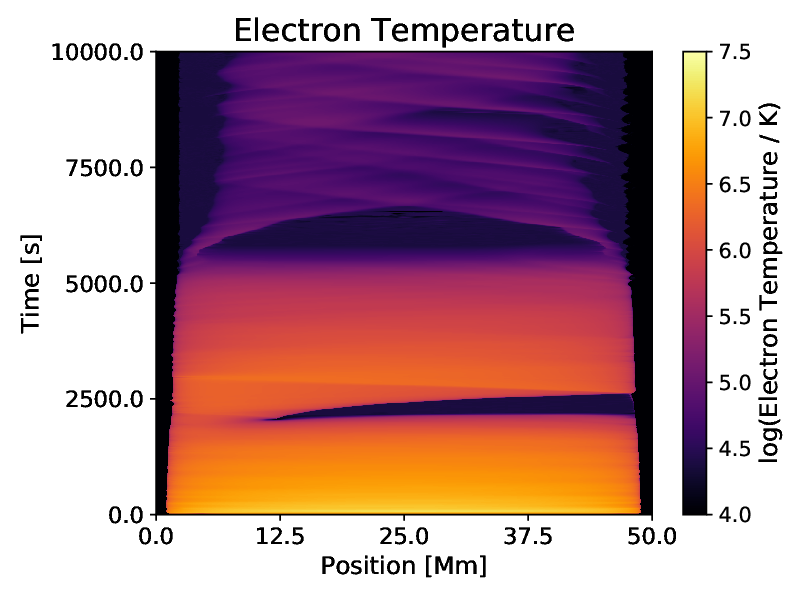}
\includegraphics[width=0.5\linewidth]{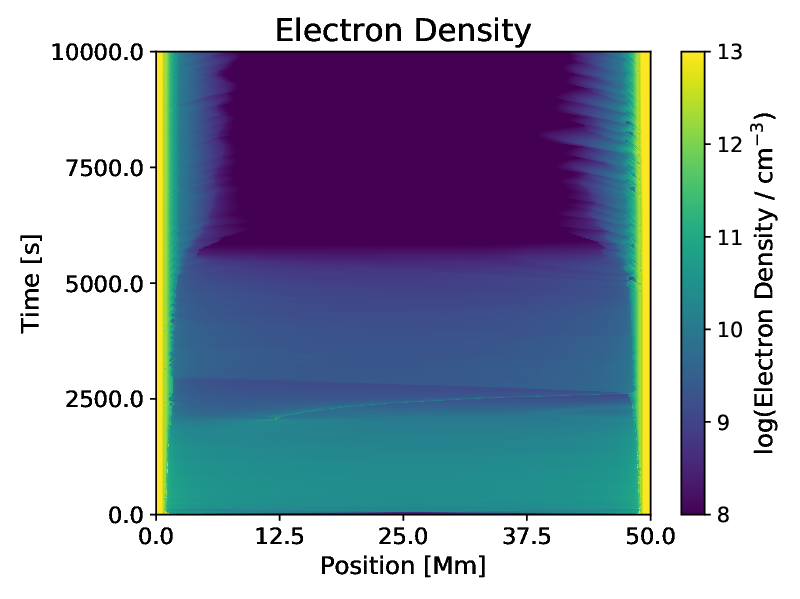}
\end{minipage}
\begin{minipage}[b]{\linewidth}
\includegraphics[width=0.5\linewidth]{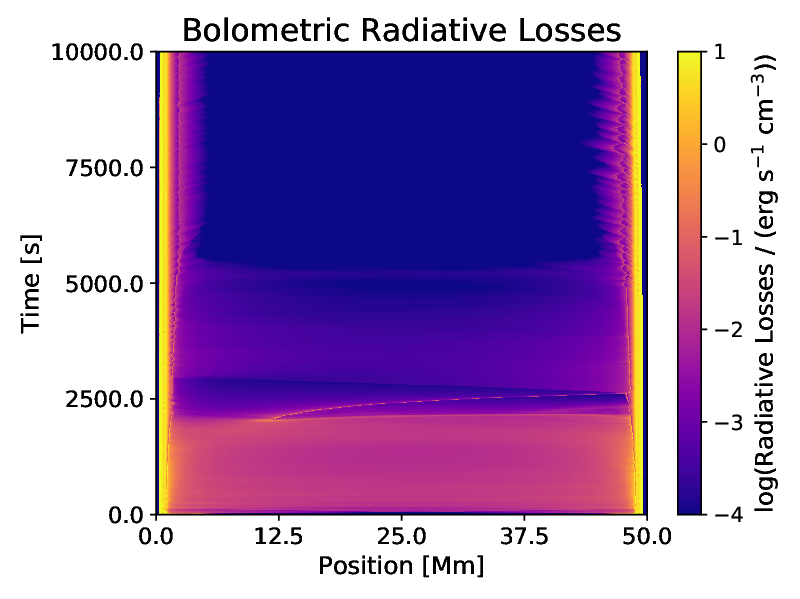}
\includegraphics[width=0.5\linewidth]{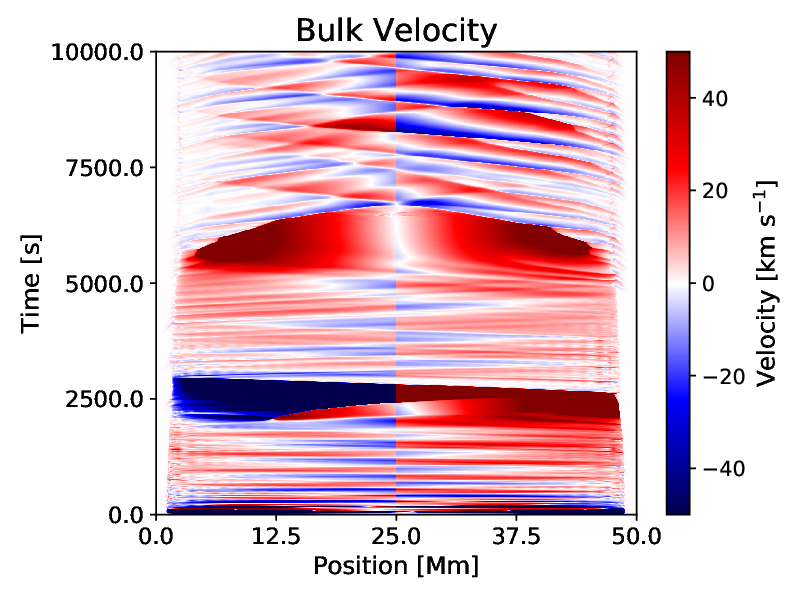}
\end{minipage}
\begin{minipage}[b]{\linewidth}
\includegraphics[width=0.5\linewidth]{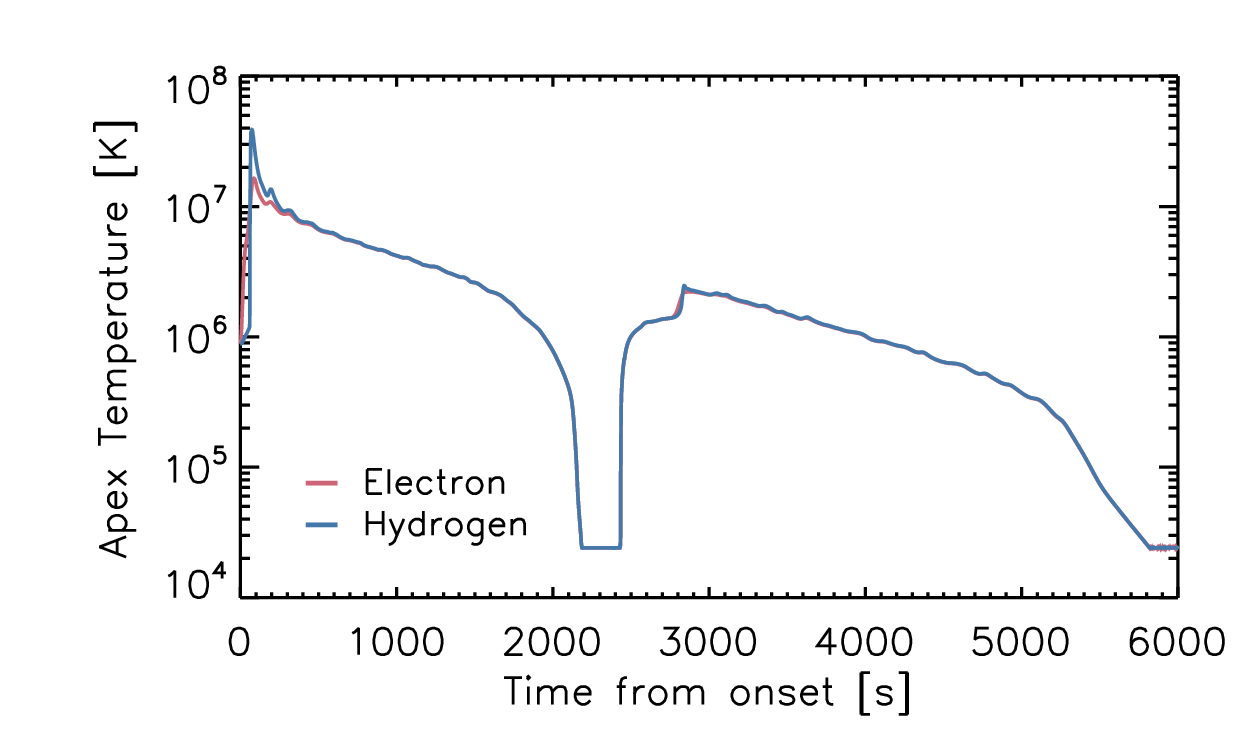}
\includegraphics[width=0.5\linewidth]{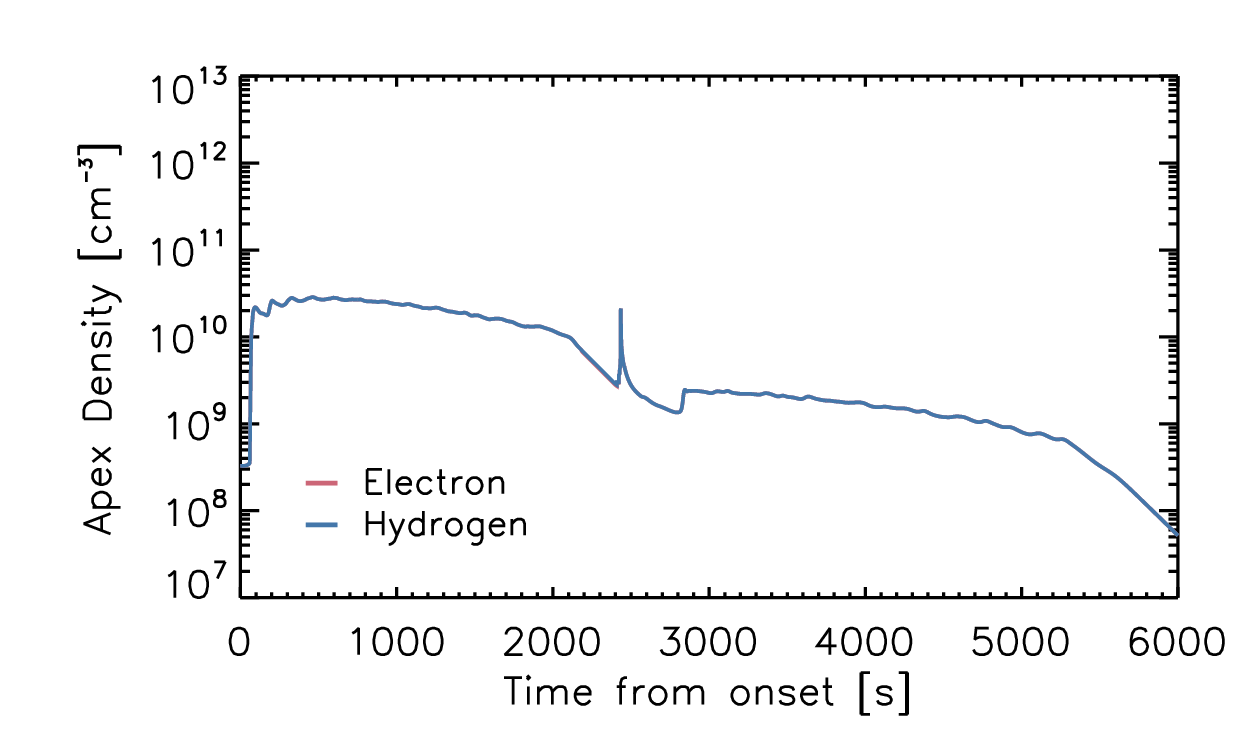}
\end{minipage}
\caption{An example of a flare rain event, similar to Figure \ref{fig:example_rain}.  The simulation combines a standard symmetric electron beam with a secondary weak footpoint heating at the left-handed footpoint.  A rain event forms that precipitates towards the right-handed footpoint.  This highlights the main point of this paper: electron beams neither produce nor inhibit the formation of coronal rain, and the formation of rain is triggered by some secondary energy transport mechanism.  What mechanism(s) can produce realistic coronal rain behavior requires further examination in the future.}
\label{fig:flare_rain}
\end{figure*}

To emphasize the point that electron beams by themselves cannot trigger the formation of coronal rain, we present an example that combines electron beam heating with a secondary, weak footpoint heating.  Figure \ref{fig:flare_rain} displays the results of such a simulation, showing the formation of a rain event.  The electron beam carries an energy flux of $10^{10}$\,erg\,s$^{-1}$\,cm$^{-2}$, with a cut-off $E_{c} = 15$ keV, spectral index $\delta = 5$, lasting for 100 s, injected symmetrically.  We have added a secondary heating event at the left-handed footpoint, lasting for 3000 s at a steady heating rate of 0.1 erg\,s$^{-1}$\,cm$^{-3}$, with a scale-height of 3 Mm, centered at a height 2 Mm above the photosphere.  The rain event forms and precipitates towards the right-handed footpoint, similarly to the quiescent rain example (Figure \ref{fig:example_rain}).  This same simulation without the secondary heating fails to produce any such rain event, behaving as the rest of the simulations in this work.  Furthermore, the formation of rain occurs in this case with other heating rates, scale heights, and temporal profiles, but the location and time at which the condensation forms seem to depend on all of these parameters.  At present, we do not diagnose what precise mechanism can be responsible for this heating, which will be an important future endeavor.  The primary point is that electron beams themselves neither cause nor inhibit the formation of rain, and that the formation of rain itself must be triggered by some secondary mechanism.  What secondary mechanism(s) and what properties of that mechanism can produce realistic coronal rain behavior requires further examination.

\begin{figure*}
\centering
\includegraphics[width=0.48\linewidth]{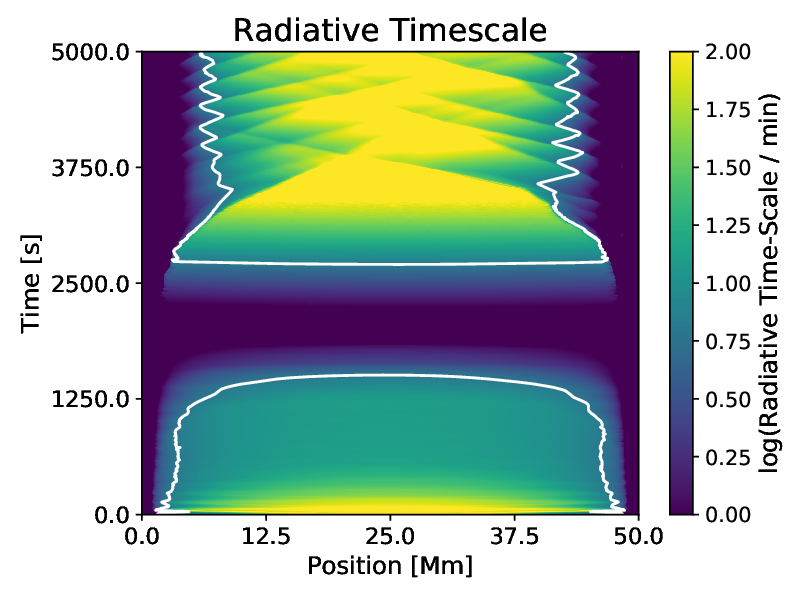}
\includegraphics[width=0.48\linewidth]{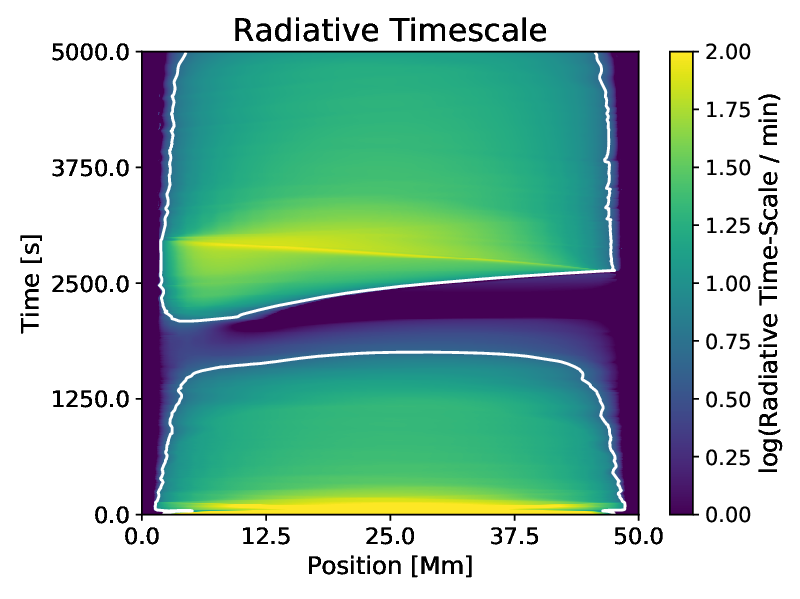}
\caption{The evolution of the radiative time-scale for the ``standard'' beam-driven flare simulation (left, corresponding to Figure \ref{fig:example_flare}) and a beam-driven simulation with secondary footpoint heating (right, corresponding to Figure \ref{fig:flare_rain}).  The white contours delineate the free-fall time for these loops (340 s).  Catastrophic cooling begins when the radiative time-scale falls below the free-fall time.  In the former case, the loop cools as a whole, while in the latter, the secondary heating perturbs the system so that a condensation forms.  \label{fig:timescale}}
\end{figure*}

In order for catastrophic cooling to occur without an additional heating term, the cooling time-scale $\tau_{c}$, which is dominated by the radiation and therefore $\propto 1/n$, must be shorter than the free-fall time $\tau_{f} = \sqrt{\frac{2 L}{\pi g_{\odot}}}$ for a semi-circular loop, where $g_{\odot}$ is the solar gravitational acceleration and $L$ the loop length.  For a loop length of 50 Mm, this corresponds to a free-fall time of around 340 s.  In the ``standard'' flare simulations in this paper, the time-scale falls below this value immediately prior to the loop collapse, but it does so across the whole length of the loop.  In cases where rain forms due to a secondary heating mechanism, this is more localized to where the condensation forms.  For example, in Figure \ref{fig:timescale}, we show the evolution of the radiative time-scale in the simulations corresponding to Figures \ref{fig:example_flare} (``standard'' beam-driven simulation, left) and \ref{fig:flare_rain} (beam with secondary footpoint heating, right).  The colors show the value of the radiative time-scale as a function of position along the loop (x-axis) and time (y-axis).  The white contours delineate the corresponding free-fall time for this loop length (340 s).  It is clear that when the time-scale falls below the free-fall time, the plasma catastrophically cools (see also \citealt{muller2003,muller2004,muller2005}).  In the former case, because the corona is approximately constant in density and temperature, the loop cools as a whole.  In the latter case, the secondary heating causes a slight gradient in temperature, so the cooling is more localized and a condensation forms.

The question then becomes what mechanism(s) can perturb this system enough to drive the formation of condensations.  There are three possible solutions to this dilemma: (1) the initial conditions of the loop are significantly denser than assumed, but this can be immediately ruled out since the flare loops would then be visible prior to the flare; (2) the cooling mechanisms, particularly the treatment of thermal conduction, needs to be modified; (3) there is some secondary mechanism that alters the dynamics.  We discuss these possibilities in turn.  

First, it is well known that the Spitzer approximation to thermal conduction \citep{spitzer1953} is not correct in high velocity regimes \citep{ljepojevic1988,ljepojevic1989}.  It is possible that a non-local contribution to the heat flux may alter the dynamics of the simulations considered here, since the tail end of the particle distribution can carry energy farther than assumed by the Spitzer approximation.  Such a non-local flux could effectively carry energy to the footpoints of the loop \citep{karpen1987}, providing an additional source of heat that can lead to thermal instabilities, which requires further examination.  It has also been suggested that the diffusion of non-thermal electrons to greater depths than predicted by the cold thick-target model is non-negligible \citep{emslie2018,jeffrey2019}.  While this diffusion would allow energy to penetrate further shortly after the onset of heating, it is still not clear that the beam itself would last long enough to induce TNE (the heating needs to be on the order of the cooling time, \citealt{johnston2019}).  

It is also suspected that turbulence can suppress thermal conduction in flaring loops \citep{bian2016a}.  This turbulent suppression can lead to additional heating in the corona by accelerated electrons, but additionally slow the conductive cooling time by a substantial amount \citep{bian2016a}.  The mean-free path of the turbulent scattering determines how effectively thermal conduction can cool the plasma \citep{bian2016b}: for long mean-free paths, the Spitzer approximation is valid, while for short mean-free paths the turbulence significantly reduces the rate of conductive cooling.  This slowed cooling allows for the loops to remain hotter and denser for longer periods of time \citep{bian2018}, which may provide a long duration heating term as the energy slowly escapes the system mostly via radiation, though the turbulent suppression of conduction is not enough on its own (without an additional heating term) to explain the long duration light curves observed in both X-ray and EUV passbands \citep{zhu2018}.  Because of its drastic effect on conduction \citep{bradshaw2019}, the role of turbulence in the formation of thermal instabilities and TNE needs to be examined in more detail.  

There are many indications that additional energy is supplied to the post-flaring loops, and that this energy may be over an order of magnitude larger than the energy released during the impulsive phase of a flare \citep{kuhar2017}.  These loops remain hot \citep{svestka1982} and dense \citep{moore1980} for longer than expected, and up-flows of material occur well into the late phase \citep{czaykowska1999}.  Further, while an electron beam could produce the heating necessary to produce the observed late phase up-flows, there would be associated HXR emission, which is not observed \citep{czaykowska2001}.  All of these issues suggest that there might be an additional heating mechanism in the late phase of solar flares.  Since this mechanism is essentially unknown, it is also unknown what impact it may have upon the formation of coronal rain in flares, but observations are clear that there is \textit{some} extra source of energy.  

One possibility that is often neglected in flare modeling is the excitation of Alfv\'enic waves.  In the Earth's magnetosphere, it is well known from \textit{in situ} measurements that magnetic reconnection excites Alfv\'en waves directly \citep{wygant2000,keiling2009}.  Magnetohydrodynamic simulations similarly predict the excitation of Alfv\'en waves \citep{birn2009,kigure2010}, as do analytic models of flare arcades \citep{tarr2017}, and there has recently been radio observations suggesting their presence in flares \citep{yu2019}.  Alfv\'en waves generated in the corona that propagate into the chromosphere can cause strong heating through their resistive dissipation \citep{reeprussell2016}, which may cause different behavior in chromospheric line evolution \citep{kerr2016}.  These waves do indeed propagate deeper into the chromosphere with time \citep{reep2018}, unlike electron beams, and are thus capable of maintaining a steady footpoint heating term.  Furthermore, at low frequencies ($\lesssim 1$\,Hz) only 10\% or less of their Poynting flux transmits through the transition region \citep{depontieu2001,reep2018}, thus allowing a slow escape of energy from the corona, which could give rise to a long duration heating.  These waves are also thought to induce a ponderomotive force that causes the first ionization potential (FIP) effect \citep{laming2015}, where elements with low FIP are over-abundant in the corona relative to their photospheric values.  Some observations suggest the FIP effect may occur in flares \citep{doschek2018}.  Future modeling efforts need to examine more closely if the dissipation of Alfv\'enic waves could be related to the formation of coronal rain.

Another possibility is compressive MHD waves associated with the reconnection event, such as fast MHD or sausage modes.  At the loop top, there is an impact from the downward reconnection jet onto the loop arcade due to propagating fast mode waves \citep{takasao2016}, which could cause compression of material that could lead to coronal condensations.  Similarly, sausage modes, which are considered one possibility to explain quasi-periodic pulsations in flares \citep{nakariakov2009,hayes2019}, are found to occur frequently in flares \citep{tian2016,nakariakov2019}.  Since they are compressive, the sausage waves could induce large enough perturbations to trigger thermal instability at the flare loop tops.  For this to happen, however, the loop would need to be over-dense and marginally stable, so that another mechanism would need to work \textit{in tandem} with these waves.  For example, the heating by an electron beam could create a situation where the loop is over-dense, but the waves themselves are the ultimate trigger for the thermal runaway.  The stability of the loop can be assessed via \textit{e.g.} the isochoric or isobaric criteria discussed by \citet{xia2011}.  These waves also need to be investigated in further detail.

We are left with a conundrum.  TNE requires extremely long duration footpoint heating, whereas flares are fundamentally impulsive, with many lasting only a few minutes, far too short a time to produce the thermal instability we expect.  This strongly suggests that there is some alternative energy transport or dynamic mechanism that is not currently being accounted for in the modeling of flares (or any impulsive event).  Future endeavors must determine what mechanism this is, and where it fits within the standard model of flares.

\leavevmode \newline

\acknowledgments  
This work was supported by NASA's \textit{Hinode} project.  \textit{Hinode} is a Japanese mission developed and launched by ISAS/JAXA with NAOJ as a domestic partner and NASA and STFC (UK) as international partners.  It is operated by these agencies in cooperation with ESA and NSC (Norway).  P.A. acknowledges funding from his STFC Ernest Rutherford Fellowship (No. ST/R004285/1).  SJB was funded for this work by the Heliophysics Supporting Research (H-SR) element of NASA ROSES (grant NNX17AD31G).  This research has made use of NASA's Astrophysics Data System.  J.W.R. and P.A. benefited from participation in the International Space Science Institute team on ``Observed Multi-Scale Variability of Coronal Loops as a Probe of Coronal Heating'' led by Clara Froment and Patrick Antolin.  The authors thank Harry Warren and Ignacio Ugarte-Urra for helpful comments in the preparation of this work and Nick Crump for help with Python.  The authors also thank the anonymous referee for constructive comments that have improved the impact and clarity of this work.
\bibliography{apj}

\begin{thebibliography}{}
\expandafter\ifx\csname natexlab\endcsname\relax\def\natexlab#1{#1}\fi
\providecommand{\url}[1]{\href{#1}{#1}}

\bibitem[{{Antiochos}(1980)}]{antiochos1980}
{Antiochos}, S.~K. 1980, \apj, 241, 385

\bibitem[{{Antiochos} \& {Klimchuk}(1991)}]{antiochos1991}
{Antiochos}, S.~K., \& {Klimchuk}, J.~A. 1991, \apj, 378, 372

\bibitem[{{Antiochos} {et~al.}(1999){Antiochos}, {MacNeice}, {Spicer}, \&
  {Klimchuk}}]{antiochos1999}
{Antiochos}, S.~K., {MacNeice}, P.~J., {Spicer}, D.~S., \& {Klimchuk}, J.~A.
  1999, \apj, 512, 985

\bibitem[{{Antolin}(2020)}]{antolin2020}
{Antolin}, P. 2020, Plasma Physics and Controlled Fusion, 62, 014016

\bibitem[{{Antolin} \& {Rouppe van der Voort}(2012)}]{antolin2012}
{Antolin}, P., \& {Rouppe van der Voort}, L. 2012, \apj, 745, 152

\bibitem[{{Auch{\`e}re} {et~al.}(2018){Auch{\`e}re}, {Froment}, {Soubri{\'e}},
  {Antolin}, {Oliver}, \& {Pelouze}}]{auchere2018}
{Auch{\`e}re}, F., {Froment}, C., {Soubri{\'e}}, E., {et~al.} 2018, \apj, 853,
  176

\bibitem[{{Bian} {et~al.}(2018){Bian}, {Emslie}, {Horne}, \&
  {Kontar}}]{bian2018}
{Bian}, N., {Emslie}, A.~G., {Horne}, D., \& {Kontar}, E.~P. 2018, \apj, 852,
  127

\bibitem[{{Bian} {et~al.}(2016{\natexlab{a}}){Bian}, {Kontar}, \&
  {Emslie}}]{bian2016a}
{Bian}, N.~H., {Kontar}, E.~P., \& {Emslie}, A.~G. 2016{\natexlab{a}}, \apj,
  824, 78

\bibitem[{{Bian} {et~al.}(2016{\natexlab{b}}){Bian}, {Watters}, {Kontar}, \&
  {Emslie}}]{bian2016b}
{Bian}, N.~H., {Watters}, J.~M., {Kontar}, E.~P., \& {Emslie}, A.~G.
  2016{\natexlab{b}}, \apj, 833, 76

\bibitem[{{Birn} {et~al.}(2009){Birn}, {Fletcher}, {Hesse}, \&
  {Neukirch}}]{birn2009}
{Birn}, J., {Fletcher}, L., {Hesse}, M., \& {Neukirch}, T. 2009, \apj, 695,
  1151

\bibitem[{{Bradshaw} \& {Cargill}(2013)}]{bradshaw2013}
{Bradshaw}, S.~J., \& {Cargill}, P.~J. 2013, \apj, 770, 12

\bibitem[{{Bradshaw} {et~al.}(2019){Bradshaw}, {Emslie}, {Bian}, \&
  {Kontar}}]{bradshaw2019}
{Bradshaw}, S.~J., {Emslie}, A.~G., {Bian}, N.~H., \& {Kontar}, E.~P. 2019,
  \apj, 880, 80

\bibitem[{{Bradshaw} \& {Mason}(2003)}]{bradshaw2003}
{Bradshaw}, S.~J., \& {Mason}, H.~E. 2003, \aap, 401, 699

\bibitem[{{Cargill}(1994)}]{cargill1994}
{Cargill}, P.~J. 1994, \apj, 422, 381

\bibitem[{{Cargill} \& {Bradshaw}(2013)}]{cargill2013}
{Cargill}, P.~J., \& {Bradshaw}, S.~J. 2013, \apj, 772, 40

\bibitem[{{Carlsson} \& {Leenaarts}(2012)}]{carlsson2012}
{Carlsson}, M., \& {Leenaarts}, J. 2012, \aap, 539, A39

\bibitem[{{Carmichael}(1964)}]{carmichael1964}
{Carmichael}, H. 1964, {A Process for Flares}, 451

\bibitem[{{Cheng} {et~al.}(1981){Cheng}, {Tandberg-Hanssen}, {Bruner}, {Orwig},
  {Frost}, {Kenny}, {Woodgate}, \& {Shine}}]{cheng1981}
{Cheng}, C.~C., {Tandberg-Hanssen}, E., {Bruner}, E.~C., {et~al.} 1981, \apj,
  248, L39

\bibitem[{{Cheng} {et~al.}(2012){Cheng}, {Qiu}, {Ding}, \& {Wang}}]{cheng2012}
{Cheng}, J.~X., {Qiu}, J., {Ding}, M.~D., \& {Wang}, H. 2012, \aap, 547, A73

\bibitem[{{Czaykowska} {et~al.}(2001){Czaykowska}, {Alexander}, \& {De
  Pontieu}}]{czaykowska2001}
{Czaykowska}, A., {Alexander}, D., \& {De Pontieu}, B. 2001, \apj, 552, 849

\bibitem[{{Czaykowska} {et~al.}(1999){Czaykowska}, {De Pontieu}, {Alexander},
  \& {Rank}}]{czaykowska1999}
{Czaykowska}, A., {De Pontieu}, B., {Alexander}, D., \& {Rank}, G. 1999, \apj,
  521, L75

\bibitem[{{De Pontieu} {et~al.}(2001){De Pontieu}, {Martens}, \&
  {Hudson}}]{depontieu2001}
{De Pontieu}, B., {Martens}, P.~C.~H., \& {Hudson}, H.~S. 2001, \apj, 558, 859

\bibitem[{{Dere} {et~al.}(2019){Dere}, {Del Zanna}, {Young}, {Landi}, \&
  {Sutherland}}]{dere2019}
{Dere}, K.~P., {Del Zanna}, G., {Young}, P.~R., {Landi}, E., \& {Sutherland},
  R.~S. 2019, \apjs, 241, 22

\bibitem[{{Doschek} {et~al.}(2018){Doschek}, {Warren}, {Harra}, {Culhane},
  {Watanabe}, \& {Hara}}]{doschek2018}
{Doschek}, G.~A., {Warren}, H.~P., {Harra}, L.~K., {et~al.} 2018, \apj, 853,
  178

\bibitem[{{Emslie}(1978)}]{emslie1978}
{Emslie}, A.~G. 1978, \apj, 224, 241

\bibitem[{{Emslie} {et~al.}(2018){Emslie}, {Bian}, \& {Kontar}}]{emslie2018}
{Emslie}, A.~G., {Bian}, N.~H., \& {Kontar}, E.~P. 2018, \apj, 862, 158

\bibitem[{{Emslie} {et~al.}(1992){Emslie}, {Li}, \& {Mariska}}]{emslie1992}
{Emslie}, A.~G., {Li}, P., \& {Mariska}, J.~T. 1992, \apj, 399, 714

\bibitem[{{Field}(1965)}]{field1965}
{Field}, G.~B. 1965, \apj, 142, 531

\bibitem[{{Fletcher} {et~al.}(2011){Fletcher}, {Dennis}, {Hudson}, {Krucker},
  {Phillips}, {Veronig}, {Battaglia}, {Bone}, {Caspi}, {Chen}, {Gallagher},
  {Grigis}, {Ji}, {Liu}, {Milligan}, \& {Temmer}}]{fletcher2011}
{Fletcher}, L., {Dennis}, B.~R., {Hudson}, H.~S., {et~al.} 2011, \ssr, 159, 19

\bibitem[{{Foukal}(1978)}]{foukal1978}
{Foukal}, P. 1978, \apj, 223, 1046

\bibitem[{{Froment} {et~al.}(2015){Froment}, {Auch{\`e}re}, {Bocchialini},
  {Buchlin}, {Guennou}, \& {Solomon}}]{froment2015}
{Froment}, C., {Auch{\`e}re}, F., {Bocchialini}, K., {et~al.} 2015, \apj, 807,
  158

\bibitem[{{Froment} {et~al.}(2018){Froment}, {Auch{\`e}re}, {Miki{\'c}},
  {Aulanier}, {Bocchialini}, {Buchlin}, {Solomon}, \&
  {Soubri{\'e}}}]{froment2018}
{Froment}, C., {Auch{\`e}re}, F., {Miki{\'c}}, Z., {et~al.} 2018, \apj, 855, 52

\bibitem[{{Hayes} {et~al.}(2019){Hayes}, {Gallagher}, {Dennis}, {Ireland},
  {Inglis}, \& {Morosan}}]{hayes2019}
{Hayes}, L.~A., {Gallagher}, P.~T., {Dennis}, B.~R., {et~al.} 2019, \apj, 875,
  33

\bibitem[{{Hirayama}(1974)}]{hirayama1974}
{Hirayama}, T. 1974, \solphys, 34, 323

\bibitem[{{Holman} {et~al.}(2003){Holman}, {Sui}, {Schwartz}, \&
  {Emslie}}]{holman2003}
{Holman}, G.~D., {Sui}, L., {Schwartz}, R.~A., \& {Emslie}, A.~G. 2003, \apj,
  595, L97

\bibitem[{{Holman} {et~al.}(2011){Holman}, {Aschwanden}, {Aurass}, {Battaglia},
  {Grigis}, {Kontar}, {Liu}, {Saint-Hilaire}, \& {Zharkova}}]{holman2011}
{Holman}, G.~D., {Aschwanden}, M.~J., {Aurass}, H., {et~al.} 2011, \ssr, 159,
  107

\bibitem[{{Jeffrey} {et~al.}(2019){Jeffrey}, {Kontar}, \&
  {Fletcher}}]{jeffrey2019}
{Jeffrey}, N. L.~S., {Kontar}, E.~P., \& {Fletcher}, L. 2019, \apj, 880, 136

\bibitem[{{Jing} {et~al.}(2016){Jing}, {Xu}, {Cao}, {Liu}, {Gary}, \&
  {Wang}}]{jing2016}
{Jing}, J., {Xu}, Y., {Cao}, W., {et~al.} 2016, Scientific Reports, 6, 24319

\bibitem[{{Johnston} {et~al.}(2019){Johnston}, {Cargill}, {Antolin}, {Hood},
  {De Moortel}, \& {Bradshaw}}]{johnston2019}
{Johnston}, C.~D., {Cargill}, P.~J., {Antolin}, P., {et~al.} 2019, \aap, 625,
  A149

\bibitem[{{Karpen} {et~al.}(2001){Karpen}, {Antiochos}, {Hohensee}, {Klimchuk},
  \& {MacNeice}}]{karpen2001}
{Karpen}, J.~T., {Antiochos}, S.~K., {Hohensee}, M., {Klimchuk}, J.~A., \&
  {MacNeice}, P.~J. 2001, \apjl, 553, L85

\bibitem[{{Karpen} \& {DeVore}(1987)}]{karpen1987}
{Karpen}, J.~T., \& {DeVore}, C.~R. 1987, \apj, 320, 904

\bibitem[{{Karpen} {et~al.}(2005){Karpen}, {Tanner}, {Antiochos}, \&
  {DeVore}}]{karpen2005}
{Karpen}, J.~T., {Tanner}, S.~E.~M., {Antiochos}, S.~K., \& {DeVore}, C.~R.
  2005, \apj, 635, 1319

\bibitem[{{Keiling}(2009)}]{keiling2009}
{Keiling}, A. 2009, \ssr, 142, 73

\bibitem[{{Kerr} {et~al.}(2016){Kerr}, {Fletcher}, {Russell}, \&
  {Allred}}]{kerr2016}
{Kerr}, G.~S., {Fletcher}, L., {Russell}, A. J.~B., \& {Allred}, J.~C. 2016,
  \apj, 827, 101

\bibitem[{{Kigure} {et~al.}(2010){Kigure}, {Takahashi}, {Shibata}, {Yokoyama},
  \& {Nozawa}}]{kigure2010}
{Kigure}, H., {Takahashi}, K., {Shibata}, K., {Yokoyama}, T., \& {Nozawa}, S.
  2010, Publications of the Astronomical Society of Japan, 62, 993

\bibitem[{{Kiplinger} {et~al.}(1983){Kiplinger}, {Dennis}, {Frost}, {Orwig}, \&
  {Emslie}}]{kiplinger1983}
{Kiplinger}, A.~L., {Dennis}, B.~R., {Frost}, K.~J., {Orwig}, L.~E., \&
  {Emslie}, A.~G. 1983, \apj, 265, L99

\bibitem[{{Klimchuk}(2019)}]{klimchuk2019b}
{Klimchuk}, J.~A. 2019, arXiv e-prints, arXiv:1911.11849

\bibitem[{{Klimchuk} \& {Luna}(2019)}]{klimchuk2019}
{Klimchuk}, J.~A., \& {Luna}, M. 2019, \apj, 884, 68

\bibitem[{{Klimchuk} {et~al.}(2008){Klimchuk}, {Patsourakos}, \&
  {Cargill}}]{klimchuk2008}
{Klimchuk}, J.~A., {Patsourakos}, S., \& {Cargill}, P.~J. 2008, \apj, 682, 1351

\bibitem[{{Kohutova} {et~al.}(2019){Kohutova}, {Verwichte}, \&
  {Froment}}]{kohutova2019}
{Kohutova}, P., {Verwichte}, E., \& {Froment}, C. 2019, \aap, 630, A123

\bibitem[{{Kopp} \& {Pneuman}(1976)}]{kopp1976}
{Kopp}, R.~A., \& {Pneuman}, G.~W. 1976, \solphys, 50, 85

\bibitem[{{Kuhar} {et~al.}(2017){Kuhar}, {Krucker}, {Hannah}, {Glesener},
  {Saint-Hilaire}, {Grefenstette}, {Hudson}, {White}, {Smith}, {Marsh},
  {Wright}, {Boggs}, {Christensen}, {Craig}, {Hailey}, {Harrison}, {Stern}, \&
  {Zhang}}]{kuhar2017}
{Kuhar}, M., {Krucker}, S., {Hannah}, I.~G., {et~al.} 2017, \apj, 835, 6

\bibitem[{{Kuin} \& {Martens}(1982)}]{kuin1982}
{Kuin}, N.~P.~M., \& {Martens}, P.~C.~H. 1982, \aap, 108, L1

\bibitem[{{Lacatus} {et~al.}(2017){Lacatus}, {Judge}, \& {Donea}}]{lacatus2017}
{Lacatus}, D.~A., {Judge}, P.~G., \& {Donea}, A. 2017, \apj, 842, 15

\bibitem[{{Laming}(2015)}]{laming2015}
{Laming}, J.~M. 2015, Living Reviews in Solar Physics, 12, 2

\bibitem[{{Lin} {et~al.}(1984){Lin}, {Schwartz}, {Kane}, {Pelling}, \&
  {Hurley}}]{lin1984}
{Lin}, R.~P., {Schwartz}, R.~A., {Kane}, S.~R., {Pelling}, R.~M., \& {Hurley},
  K.~C. 1984, \apj, 283, 421

\bibitem[{{Ljepojevic} \& {MacNeice}(1988)}]{ljepojevic1988}
{Ljepojevic}, N.~N., \& {MacNeice}, P. 1988, \solphys, 117, 123

\bibitem[{{Ljepojevic} \& {MacNeice}(1989)}]{ljepojevic1989}
---. 1989, Physical Review A, 40, 981

\bibitem[{{Miki{\'c}} {et~al.}(2013){Miki{\'c}}, {Lionello}, {Mok}, {Linker},
  \& {Winebarger}}]{mikic2013}
{Miki{\'c}}, Z., {Lionello}, R., {Mok}, Y., {Linker}, J.~A., \& {Winebarger},
  A.~R. 2013, \apj, 773, 94

\bibitem[{{Moore} {et~al.}(1980){Moore}, {McKenzie}, {Svestka}, {Widing},
  {Dere}, {Antiochos}, {Dodson-Prince}, {Hiei}, {Krall}, \&
  {Krieger}}]{moore1980}
{Moore}, R., {McKenzie}, D.~L., {Svestka}, Z., {et~al.} 1980, in Skylab Solar
  Workshop II, ed. P.~A. {Sturrock}, 341--409

\bibitem[{{Moravec} {et~al.}(2016){Moravec}, {Varady}, {Ka{\v{s}}parov{\'a}},
  \& {Kramoli{\v{s}}}}]{moravec2016}
{Moravec}, Z., {Varady}, M., {Ka{\v{s}}parov{\'a}}, J., \& {Kramoli{\v{s}}}, D.
  2016, Astronomische Nachrichten, 337, 1020

\bibitem[{{M{\"u}ller} {et~al.}(2005){M{\"u}ller}, {De Groof}, {Hansteen}, \&
  {Peter}}]{muller2005}
{M{\"u}ller}, D.~A.~N., {De Groof}, A., {Hansteen}, V.~H., \& {Peter}, H. 2005,
  \aap, 436, 1067

\bibitem[{{M{\"u}ller} {et~al.}(2003){M{\"u}ller}, {Hansteen}, \&
  {Peter}}]{muller2003}
{M{\"u}ller}, D.~A.~N., {Hansteen}, V.~H., \& {Peter}, H. 2003, \aap, 411, 605

\bibitem[{{M{\"u}ller} {et~al.}(2004){M{\"u}ller}, {Peter}, \&
  {Hansteen}}]{muller2004}
{M{\"u}ller}, D.~A.~N., {Peter}, H., \& {Hansteen}, V.~H. 2004, \aap, 424, 289

\bibitem[{{Nagai} \& {Emslie}(1984)}]{nagai1984}
{Nagai}, F., \& {Emslie}, A.~G. 1984, \apj, 279, 896

\bibitem[{{Nakariakov} {et~al.}(2019){Nakariakov}, {Kolotkov}, {Kupriyanova},
  {Mehta}, {Pugh}, {Lee}, \& {Broomhall}}]{nakariakov2019}
{Nakariakov}, V.~M., {Kolotkov}, D.~Y., {Kupriyanova}, E.~G., {et~al.} 2019,
  Plasma Physics and Controlled Fusion, 61, 014024

\bibitem[{{Nakariakov} \& {Melnikov}(2009)}]{nakariakov2009}
{Nakariakov}, V.~M., \& {Melnikov}, V.~F. 2009, \ssr, 149, 119

\bibitem[{{Oliver} {et~al.}(2014){Oliver}, {Soler}, {Terradas}, {Zaqarashvili},
  \& {Khodachenko}}]{oliver2014}
{Oliver}, R., {Soler}, R., {Terradas}, J., {Zaqarashvili}, T.~V., \&
  {Khodachenko}, M.~L. 2014, \apj, 784, 21

\bibitem[{{Parker}(1953)}]{parker1953}
{Parker}, E.~N. 1953, \apj, 117, 431

\bibitem[{{Reep} {et~al.}(2015){Reep}, {Bradshaw}, \& {Alexander}}]{reep2015}
{Reep}, J.~W., {Bradshaw}, S.~J., \& {Alexander}, D. 2015, \apj, 808, 177

\bibitem[{{Reep} {et~al.}(2019){Reep}, {Bradshaw}, {Crump}, \&
  {Warren}}]{reep2019}
{Reep}, J.~W., {Bradshaw}, S.~J., {Crump}, N.~A., \& {Warren}, H.~P. 2019,
  \apj, 871, 18

\bibitem[{{Reep} {et~al.}(2016){Reep}, {Bradshaw}, \& {Holman}}]{reep2016}
{Reep}, J.~W., {Bradshaw}, S.~J., \& {Holman}, G.~D. 2016, \apj, 818, 44

\bibitem[{{Reep} {et~al.}(2013){Reep}, {Bradshaw}, \& {McAteer}}]{reep2013}
{Reep}, J.~W., {Bradshaw}, S.~J., \& {McAteer}, R.~T.~J. 2013, \apj, 778, 76

\bibitem[{{Reep} \& {Knizhnik}(2019)}]{reepknizhnik2019}
{Reep}, J.~W., \& {Knizhnik}, K.~J. 2019, \apj, 874, 157

\bibitem[{{Reep} {et~al.}(2018{\natexlab{a}}){Reep}, {Polito}, {Warren}, \&
  {Crump}}]{reeppolito2018}
{Reep}, J.~W., {Polito}, V., {Warren}, H.~P., \& {Crump}, N.~A.
  2018{\natexlab{a}}, \apj, 856, 149

\bibitem[{{Reep} \& {Russell}(2016)}]{reeprussell2016}
{Reep}, J.~W., \& {Russell}, A.~J.~B. 2016, \apjl, 818, L20

\bibitem[{{Reep} {et~al.}(2018{\natexlab{b}}){Reep}, {Russell}, {Tarr}, \&
  {Leake}}]{reep2018}
{Reep}, J.~W., {Russell}, A. J.~B., {Tarr}, L.~A., \& {Leake}, J.~E.
  2018{\natexlab{b}}, \apj, 853, 101

\bibitem[{{Rosner} {et~al.}(1978){Rosner}, {Tucker}, \& {Vaiana}}]{rosner1978}
{Rosner}, R., {Tucker}, W.~H., \& {Vaiana}, G.~S. 1978, \apj, 220, 643

\bibitem[{{Scullion} {et~al.}(2016){Scullion}, {Rouppe van der Voort},
  {Antolin}, {Wedemeyer}, {Vissers}, {Kontar}, \& {Gallagher}}]{scullion2016}
{Scullion}, E., {Rouppe van der Voort}, L., {Antolin}, P., {et~al.} 2016, \apj,
  833, 184

\bibitem[{{Spitzer} \& {H{\"a}rm}(1953)}]{spitzer1953}
{Spitzer}, L., \& {H{\"a}rm}, R. 1953, Physical Review, 89, 977

\bibitem[{{Sturrock}(1966)}]{sturrock1966}
{Sturrock}, P.~A. 1966, \nat, 211, 695

\bibitem[{{Takasao} \& {Shibata}(2016)}]{takasao2016}
{Takasao}, S., \& {Shibata}, K. 2016, \apj, 823, 150

\bibitem[{{Tarr}(2017)}]{tarr2017}
{Tarr}, L.~A. 2017, \apj, 847, 1

\bibitem[{{Tian} {et~al.}(2016){Tian}, {Young}, {Reeves}, {Wang}, {Antolin},
  {Chen}, \& {He}}]{tian2016}
{Tian}, H., {Young}, P.~R., {Reeves}, K.~K., {et~al.} 2016, \apjl, 823, L16

\bibitem[{{Vernazza} {et~al.}(1981){Vernazza}, {Avrett}, \&
  {Loeser}}]{vernazza1981}
{Vernazza}, J.~E., {Avrett}, E.~H., \& {Loeser}, R. 1981, \apjs, 45, 635

\bibitem[{{{\v{S}}vestka} {et~al.}(1982){{\v{S}}vestka}, {Dodson-Prince},
  {Martin}, {Mohler}, {Moore}, {Nolte}, \& {Petrasso}}]{svestka1982}
{{\v{S}}vestka}, Z., {Dodson-Prince}, H.~W., {Martin}, S.~F., {et~al.} 1982,
  \solphys, 78, 271

\bibitem[{{Warren}(2006)}]{warren2006}
{Warren}, H.~P. 2006, \apj, 637, 522

\bibitem[{{Warren} {et~al.}(2018){Warren}, {Brooks}, {Ugarte-Urra}, {Reep},
  {Crump}, \& {Doschek}}]{warren2018}
{Warren}, H.~P., {Brooks}, D.~H., {Ugarte-Urra}, I., {et~al.} 2018, \apj, 854,
  122

\bibitem[{{Warren} {et~al.}(2016){Warren}, {Reep}, {Crump}, \&
  {Sim{\~o}es}}]{warren2016}
{Warren}, H.~P., {Reep}, J.~W., {Crump}, N.~A., \& {Sim{\~o}es}, P. J.~A. 2016,
  \apj, 829, 35

\bibitem[{{Warren} \& {Warshall}(2001)}]{warren2001}
{Warren}, H.~P., \& {Warshall}, A.~D. 2001, \apj, 560, L87

\bibitem[{{Wygant} {et~al.}(2000){Wygant}, {Keiling}, {Cattell}, {Johnson},
  {Lysak}, {Temerin}, {Mozer}, {Kletzing}, {Scudder}, {Peterson}, {Russell},
  {Parks}, {Brittnacher}, {Germany}, \& {Spann}}]{wygant2000}
{Wygant}, J.~R., {Keiling}, A., {Cattell}, C.~A., {et~al.} 2000, Journal of
  Geophysical Research, 105, 18,675

\bibitem[{{Xia} {et~al.}(2011){Xia}, {Chen}, {Keppens}, \& {van
  Marle}}]{xia2011}
{Xia}, C., {Chen}, P.~F., {Keppens}, R., \& {van Marle}, A.~J. 2011, \apj, 737,
  27

\bibitem[{{Yu} \& {Chen}(2019)}]{yu2019}
{Yu}, S., \& {Chen}, B. 2019, \apj, 872, 71

\bibitem[{{Zhu} {et~al.}(2018){Zhu}, {Qiu}, \& {Longcope}}]{zhu2018}
{Zhu}, C., {Qiu}, J., \& {Longcope}, D.~W. 2018, \apj, 856, 27

\end{thebibliography}
\bibliographystyle{aasjournal}

\end{document}